\documentclass[structabstract]{aa}  
\def\HH        {H$_{2}$}

\def\meth      {CH$_{3}$OH}
\def\thmeth    {$^{13}$CH$_{3}$OH}
\def\dmeth     {CH$_{2}$DOH}
\def\methd     {CH$_{3}$OD}

\def\mf        {HCOOCH$_{3}$}
\def\water     {H$_2$O}

\def\kms       {km~s$^{-1}$}
\def\cmm       {cm$^{-2}$}

\def\vlsr      {$V_{\rm LSR}$}

\def\jb        {Jy~beam$^{-1}$}
\def\mjb       {mJy~beam$^{-1}$}

\usepackage{graphicx}
\usepackage{lscape}
\usepackage{natbib}
\usepackage{txfonts}
\begin{document}
   \title{Deuterated methanol in Orion BN/KL
   \thanks{Based on observations carried out with the IRAM Plateau de Bure Interferometer. IRAM is supported by INSU/CNRS (France), MPG (Germany), and IGN (Spain).}}
%   \subtitle{I. Overviewing the $\kappa$-mechanism}

   \author{T.-C. Peng \inst{1,2}
          \and D. Despois\inst{1,2}
          \and N. Brouillet\inst{1,2}
          \and B. Parise\inst{3}
          \and A. Baudry\inst{1,2}
%          \fnmsep\thanks{Just to show the usage
 %         of the elements in the author field}
          }

   \institute{Univ. Bordeaux, LAB, UMR 5804, F-33270, Floirac, France \\
      \email{Tzu-Cheng.Peng@obs.u-bordeaux1.fr}
\and CNRS, LAB, UMR 5804, F-33270, Floirac, France
              \and  Max-Planck-Institut f\"ur Radioastronomie (MPIfR),
              Auf dem H\"ugel 69, 53121 Bonn, Germany
%	             \thanks{The university of heaven temporarily does not
%	                     accept e-mails}
             }

   \date{}

% \abstract{}{}{}{}{} 
% 5 {} token are mandatory
 
  \abstract
   {}
   {Deuterated molecules have been detected and studied toward Orion BN/KL in the past decades, mostly with single-dish telescopes. However, high angular resolution data are critical not only for interpreting the spatial distribution of the deuteration ratio but also for understanding this complex region in terms of cloud evolution involving star-forming activities and stellar feedbacks. Therefore, it is important to investigate the deuterated ratio of methanol, one of the most abundant grain-surface species, on a scale of a few arcseconds to better understand the physical conditions related to deuteration in Orion BN/KL.}
   {Orion BN/KL was extensively observed with the IRAM Plateau de Bure Interferometer from 1999 to 2007 in the 1 to 3 mm range. The angular resolution varies from $1\farcs8\times0\farcs8$ to $3\farcs6\times2\farcs3$ and the spectral resolution varies from 0.4 to 1.9 \kms. Deuterated methanol \dmeth\ and \methd\ and \meth\ lines were searched for within our 3 mm and 1.3 mm data sets.}
   {We present here the first high angular resolution ($1\farcs8\times0\farcs8$) images of deuterated methanol \dmeth\ in Orion BN/KL. Six \dmeth\ lines were detected around 105.8, 223.5, and 225.9 GHz. In addition, three E-type methanol lines around 101--102 GHz were detected and were used to derive the corresponding \meth\ rotational temperatures and column densities toward different regions across Orion BN/KL. The strongest \dmeth\ and \meth\ emissions come from the Hot Core southwest region with a velocity that is typical of the Compact Ridge (\vlsr\ $\approx8$ \kms). We derive [\dmeth]/[\meth] abundance ratios of $0.8-1.3\times10^{-3}$ toward three \dmeth\ emission peaks. A new transition of \methd\ was detected at 226.2 GHz for the first time in the interstellar medium. Its distribution is similar to that of \dmeth. Besides, we find that the [\dmeth]/[\methd] abundance ratios are lower than unity in the central part of BN/KL. Furthermore, the HDO $3_{1,2}-2_{2,1}$ line at 225.9 GHz was detected and its emission distribution shows a shift of a few arcseconds with respect to the deuterated methanol emission that likely results from different excitation effects. The deuteration ratios derived along Orion BN/KL are not markedly different from one clump to another. However, various processes such as slow heating due to ongoing star formation, heating by luminous infrared sources, or heating by shocks could be competing to explain some local differences observed for these ratios.}
   {}

   \maketitle
%
%________________________________________________________________

\section{Introduction}

\begin{table*}
\caption{Observational parameters of the PdBI data sets}             % title of Table
\label{table-data}      % is used to refer this table in the text
\centering                          % used for centering table
\begin{tabular}{lcccccccc}        % centered columns (4 columns)
\hline\hline                 % inserts double horizontal lines
Bandwidth  & Observation date & Configuration & Flux conversion    & RMS noise         & $\theta_{\rm HPBW}$\tablefootmark{a} & $\delta{\rm v}$\tablefootmark{b}  & \multicolumn{2}{c}{$\theta_{\rm syn}$\tablefootmark{c}}       \\    
(GHz)      &                  &               & (1 \jb) & (\mjb) & (\arcsec)           & (\kms)            & (\arcsec$\times$\arcsec) & PA (\degr)              \\
       
\hline                        % inserts single horizontal line 

101.178--101.717  &  2003--2006   &  BC  & 15.8 K  &  2.9 &  49.7  & 1.85  & $3.79\times1.99$  & 22 \\
105.655--105.726  &  2005 Aug--Nov  &  D   &  2.9 K  &  7.6 &  47.7  & 0.89  & $7.13\times5.36$  &  9 \\
223.402--223.941  &  2003--2007   &  BC  & 17.3 K  & 21.2 &  22.5  & 0.84  & $1.79\times0.79$  & 14 \\
225.805--225.942  &  2005 Sep--Nov  &  D   &  2.9 K  & 40.8 &  22.3  & 0.42  & $3.63\times2.26$  & 12 \\
225.990--226.192  &  2005 Sep--Nov  &  D   &  2.9 K  & 40.8 &  22.3  & 0.42  & $3.63\times2.25$  & 12 \\

\hline                                   %inserts single line
\end{tabular}
\tablefoot{
\tablefoottext{a}{Primary beam size}
\tablefoottext{b}{Channel separation}
\tablefoottext{c}{Synthesized beam size}
}

\end{table*}

Deuterium chemistry in the interstellar medium (ISM) has been intensively studied in recent decades. More and more deuterated molecules have been found as well as multiply-deuterated species, e.g., ND$_{3}$ \citep{van der Tak2002,Lis2002} and CD$_3$OH \citep{Parise2004}. Deuterium chemistry models have also been revised to explain the observations that show a strong enhancement of deuterium-bearing molecule abundances in star-forming regions \citep[e.g.,][]{Roberts2003,Roberts2000,Charnley1997}, compared with the D/H ratio of $2-3\times10^{-5}$ in the local interstellar gas \citep[see, e.g.,][and references therein]{Linsky2006}. Those strong enhancements (by a factor of a few thousand) are seen mostly in methanol, ammonia, water, and formaldehyde, leading to the abundance ratios $>0.1$ compared with their non-deuterated analogs. The formation of these molecules can be largely explained by grain-surface reactions that provide a natural explanation for the abundant doubly or multiply deuterated molecules that ion-molecule chemistry failed to predict \citep{Turner1990}.

Many deuterated molecules were first detected toward the Orion Becklin-Neugebauer/Kleinmann-Low (BN/KL) region \citep{Becklin1967,Kleinmann1967}, one of the closest \citep[$414\pm7$ pc,][]{Menten2007} and most-studied star-forming regions in the sky. For example, deuterated water HDO and deuterated ammonia NH$_2$D were first detected by \citet{Turner1975} and \citet{Rodriguez1978}, and the single-deuterated methanol molecules \methd\ and \dmeth\ were first detected toward the same region by \citet{Mauersberger1988} and \citet{Jacq1993}, respectively. Although grain-surface chemical models can well explain the high abundance of those deuterated molecules in molecular clouds, the deuteration branching ratios in the same species predicted by models disagree with observations. For instance, \citet{Charnley1997} showed that the formation of deuterated methanol on grains based on the addition of H and D atoms to CO always leads to [\dmeth]/[\methd] abundance ratios of about 3. This prediction is in conflict with the [\dmeth]/[\methd] ratio of 1.1--1.5 observed by \citet{Jacq1993} in Orion BN/KL. However, \citet{Rodgers2002} later pointed out that without taking into account other surface species such as CO and H$_2$CO and possibly different energy barriers involved in the reaction scheme, the [\dmeth]/[\methd] ratio of 3 is an artificial result of the model assumptions. In addition, most proposed models that include surface chemistry strongly depend on local environment where temperature and gas density play important roles in the chemical reaction rates. This is especially true for the Orion BN/KL region where contributions from stellar feedbacks (e.g., ultraviolet photons) and star formation activities (e.g., outflows/shocks) alter the warm-up history of the cloud, involving both grain surface and gas-phase chemistry. Hence, it is crucial to investigate the BN/KL deuterated methanol distribution with high spatial resolutions so that the physical conditions of individual clumps are properly constrained and the deuteration ratios can be related to specific physical processes. Additionally, deeper insight into the processes at play can be gained by comparing \dmeth\ maps with another major deuterated species, HDO.

The massive star-forming region Orion BN/KL is complicated not only because of the interactions between outflows and the ambient material, but also because of its rich and complex chemistry at the so-called Orion Hot Core and Compact Ridge regions \citep[see, e.g.,][]{Blake1987}. Owing to larger single-dish beam sizes in the early spectral line studies of Orion BN/KL, molecular line profiles were usually decomposed into several components according to their local standard of rest (LSR) velocities and line widths. The Orion hot molecular core (Hot Core) is usually characterized by its velocity component at \vlsr=5--6 \kms\ and a broad line width of about 5--10 \kms. It has been identified in the interferometric maps as a strong mm/submm continuum emission peak close to the infrared (IR) source IRc2 \citep[see, e.g.,][]{Gezari1998,Downes1981,Rieke1973}, whereas many molecular emission peaks observed around 8 \kms\ are displaced from the Hot Core by about 4\arcsec\ to the southwest \citep[Hot Core Southwest, HC-SW; see, e.g., works of][and references therein]{Favre2011,Wang2010,Tang2010,Friedel2008}.

The 8--9 \kms\ LSR velocity component of the Orion Compact Ridge exhibits a relatively narrow line width (3--5 \kms). The exact location of the Orion Compact Ridge is somewhat ambiguous, but recent interferometric observations \citep[e.g.,][]{Favre2011,Friedel2008} suggest that it is located 10\arcsec--15\arcsec\ to the southwest of the Hot Core (the strongest dust continuum peak). It is important to mention that the Orion Hot Core is located within the NE-SW dense ridge of the BN/KL region seen in dust continuum emission \citep[][]{Favre2011,Tang2010}, and is part of the hierarchical filamentary structure seen on a larger scale in OMC-1 \citep[e.g., mid-$J$ CO images by][]{Peng2012}. Additionally, the Orion Compact Ridge is located at the southern part of this dense ridge, the bottom part of the V-shaped region seen in many molecular lines, e.g., CS \citep{Chandler1997}, SO \citep{Wright1996}, HCOOCH$_3$ \citep{Favre2011}, and NH$_3$ \citep{Goddi2011}. 

The main goal of this paper is to investigate the deuteration ratios in Orion BN/KL, and address the possible causes for the abundant deuterated methanol in this region. We present the first high angular resolution ($1\farcs8\times0\farcs8$) images of \dmeth\ toward the Orion BN/KL region (\S\ref{dmeth-result}). \meth\ maps were also obtained from the same data sets with a $3\farcs8\times2\farcs0$ resolution (\S\ref{meth-result}). The \methd\ map and HDO result are shown in \S\ref{methd-result} and \S\ref{hdo-result}, respectively. We discuss methanol deuteration in \S\ref{meth-diss} for \dmeth\ and \S\ref{methd-diss} for \methd. In \S\ref{Herschel-compare}, our own \meth\ data are discussed in the light of the spectral line profiles obtained at much higher frequencies with {\it Herschel}. Comparison of our deuterated methanol and methanol maps with our deuterated water maps is presented in \S\ref{HDO-compare}, and water and methanol deuteration ratios are discussed in \S\ref{deuteration-ratio}.

%________________________________________________________________
\section{Observations and spectroscopy}

\subsection{IRAM observations}

The data used in this study are part of the large 1--3 mm data sets obtained from 1999 to 2007 using PdBI\footnote{The IRAM Plateau de Bure Interferometer. IRAM is supported by INSU/CNRS (France), MPG (Germany) and IGN (Spain).} toward the Orion BN/KL region \citep[see][for more observational details]{Favre2011}. Four data sets (Table \ref{table-data}) and five antennas equipped with two SIS receivers were used in this study. The quasars 0458--020, 0528+134, 0605--085, and 0607--157, and the BL Lac source 0420--014 were observed for phase and amplitude calibration. The six units of the correlator allowed us to achieve different bandwidths and spectral resolutions. Using the 30m single-dish data (J. Cernicharo, priv. comm.), the missing flux for the \dmeth\ lines around 223 GHz is estimated to be 20\%--50\%. The large uncertainty in this estimate is due to line confusion and the difficulty in determining spectral baselines in the 30m data. The missing flux in our \meth\ line interferometric observations around 101 GHz is estimated to be $\lesssim6\%$. As for the HDO, \dmeth, and \methd\ data at 226 GHz, we compared our PdBI HDO line emission at 225896.7 MHz with the 30m data of \citet{Jacq1990}, and we find that the missing flux for this HDO line is about 23\% toward IRc2. Since the \dmeth\ and \methd\ emissions are less extended than the HDO emission, the missing fluxes of \dmeth\ and \methd\ are likely less than 23\% at 226 GHz.

The PdBI data were reduced with the GILDAS\footnote{http://www.iram.fr/IRAMFR/GILDAS/} package, and the continuum emission was subtracted by selecting line-free channels. Our continuum emission images were presented in \citet{Favre2011} where the \HH\ column densities of selected clumps were also estimated. The data cube was then cleaned channel-by-channel with the Clark algorithm \citep{Clark1980} implemented in the GILDAS package.

\subsection{\dmeth\ spectroscopy}

\begin{table}
\caption{Spectroscopic parameters of the detected lines}             % title of Table
\label{table-CH2DOH-CH3OH}      % is used to refer this table in the text
\centering                          % used for centering table
\begin{tabular}{lccc}        % centered columns (4 columns)
\hline\hline                 % inserts double horizontal lines
Frequency\tablefootmark{a}  & Transition             & $E_{\rm up}/k$ & $S\mu^2$  \\    
(MHz)      & ($J_{k_{a},k_{c}}$)    & (K)          &  (D$^2$)   \\

\hline 
\multicolumn{4}{c}{\dmeth} \\
\hline                        % inserts single horizontal line

105806.4100 & $11_{1,10}-11_{0,11}$ $e_1$                    & 159.4 & 6.11  \\     
223422.2629 &     $5_{2,4}-4_{2,3}$ $e_0$                    &  48.4 & 5.15  \\     
223616.1420 &     $5_{4,2}-4_{4,1}$, $5_{4,1}-4_{4,0}$ $e_0$ &  95.2 & 2.15\tablefootmark{b}  \\  
223691.5380 &     $5_{3,3}-4_{3,2}$ $e_0$                    &  68.2 & 3.96  \\     
223697.1880 &     $5_{3,2}-4_{3,1}$ $e_0$                    &  68.2 & 3.96  \\  
225878.2540 &     $3_{1,3}-2_{0,2}$ $o_0$                    &  35.7 & 2.28  \\  
\hline 
\multicolumn{4}{c}{\meth} \\
\hline 
101185.4530 & $6_{-2}-6_{1}$ E &  74.7 & 0.021  \\     
101293.4150 & $7_{-2}-7_{1}$ E &  91.0 & 0.046  \\     
101469.8050 & $8_{-2}-8_{1}$ E & 109.6 & 0.091  \\  
\hline 
\multicolumn{4}{c}{\methd} \\
\hline
226185.9300 & $5_{-1}-4_{-1}$ E &  37.3 & 3.4  \\  
\hline 
\multicolumn{4}{c}{HDO} \\
\hline 
225896.7200 & $3_{1,2}-2_{2,1}$  & 167.6 & 0.69  \\  

\hline                                   %inserts single line
\end{tabular}
\tablefoot{
\tablefoottext{a}{Line frequencies from this work}
\tablefoottext{b}{Line strength of each single line}
}

\end{table}

\dmeth\ is an asymmetric top molecule similar to \meth\ where one H atom is replaced by a D atom in the methyl group (--CH$_3$). Therefore, the threefold symmetry (C$_{3\rm V}$) of the methyl group in the internal rotation of \meth\ is broken, and the \meth\ symmetry states A, E1, and E2 become $e_0$, $e_1$, and $o_1$ in \dmeth, which are more difficult to deal with theoretically and experimentally \citep[see, e.g.,][]{Lauvergnat2009,Mukhopadhyay2002}, leading to only few published studies of accurate line frequency measurements to date. The \dmeth\ substate symmetry can be characterized as even ($e$) or odd ($o$) according to the plane of symmetry defined by the DCO or COH plane. For example, the minimum potential energy for \dmeth\ occurs in the ``trans'' substate where the D atom is opposite to OH, and the secondary minimum occurs at the ``gauche'' positions where the D atom is located at --120 degrees from the ``trans'' position \citep[][]{Mukhopadhyay2002}. Therefore, the ground torsional states of \dmeth\ contain three substates, $e_0$ (``trans''), $e_1$ (symmetric ``gauche''), and $o_1$ (antisymmetric ``gauche'') with an increasing energy, similar to ethanol \citep[see, e.g.,][]{Pearson1997,Mukhopadhyay2002}. The selection rules for the same symmetry states (e.g., $e_0-e_0$) allow a- and b-type transitions, and c-type transitions are allowed for involving different symmetry states (e.g., $o_1-e_1$) in \dmeth. The \dmeth\ molecular parameters are listed in Table \ref{table-CH2DOH-CH3OH}. The \dmeth\ molecular parameters were provided by one of us (B. Parise, see Appendix \ref{app1}), and the \meth\ molecular parameters were taken from the JPL database\footnote{http://spec.jpl.nasa.gov/} \citep[see calculations of][]{Xu2008}.

       \begin{figure*}
   \centering
   \includegraphics[angle=-90,width=0.95\textwidth]{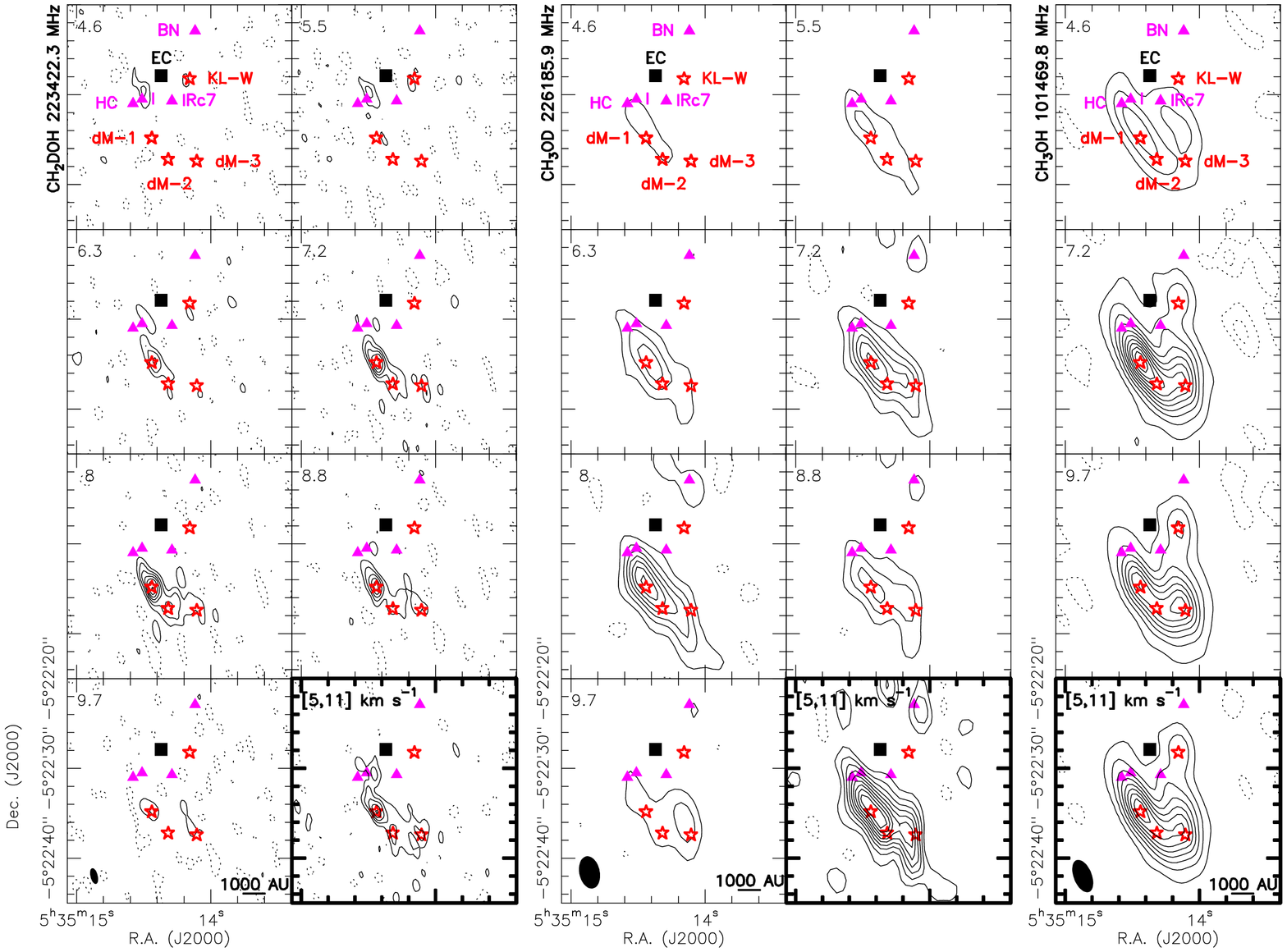}
   \caption{Left: Channel maps of the \dmeth\ $5_{2,4}-4_{2,3}$ ($E_{\rm up}/k=95.2$ K) emission at 223422.3 MHz for a synthesized beam of $1\farcs8\times0\farcs8$. Contours run from 40 \mjb\ (2 $\sigma$) to 280 \mjb\ (4.84 K) in steps of 40 \mjb, and the dashed contours represent --20 \mjb. The bottom-right panel shows the integrated intensity (from 5 to 11 \kms) in contours running from 0.2 to 1.0 \jb\ \kms\ in steps of 0.2 \jb\ \kms, and the dashed contours represent --0.08 \jb\ \kms. Middle: Channel maps of the \methd\ emission at 226185.9 MHz ($E_{\rm up}/k=37.3$ K) for a synthesized beam of $3\farcs6\times2\farcs3$. Contours run from 0.12 to 2.52 \jb\ in steps of 0.04 \jb\ (3 $\sigma$), and the dashed contours represent --0.08 \jb. The bottom-right panel shows the integrated intensity of \methd\ (from 5 to 11 \kms) in contours running from 10\% to 90\% in step of 10\% of the peak intensity (3.1 \jb\ \kms), and the dashed contours represent --10\% of the peak intensity. Right: Channel maps of the \meth\ $8_{-2}-8_{1}$ E ($E_{\rm up}/k=109.6$ K) emission at 101469.8 MHz for a synthesized beam of $3\farcs8\times2\farcs0$. Contours run from 40 to 490 \mjb\ (7.74 K) in steps of 50 \mjb. The dashed contours represent --30 (1 $\sigma$) and --90 \mjb. The bottom panel shows the integrated intensity of \meth\ (from 5 to 11 \kms) in contours running from 0.3 to 3.0 \jb\ \kms\ in steps of 0.3 \jb\ \kms, and the dashed contours represent --0.03 \jb\ \kms. The black square marks the center of explosion according to \citet{Zapata2009}. The positions of source BN, source $I$, the Hot Core (HC), and IRc7 are marked by magenta triangles. The positions of deuterated methanol emission peaks (dM-1, dM-2, and dM-3) and KL-W are marked by red stars. Note that the channel separation in the \methd\ channel maps are resampled and only the three selected \meth\ channel maps are shown here for clarity (see Fig. \ref{Fig-CH2DOH-CH3OH-chmap} for the full \meth\ channel maps).}
              \label{Fig-CH2DOH-CH3OD-CH3OH-chmap}
    \end{figure*}

%________________________________________________________________
\section{Results}

\subsection{\dmeth\label{dmeth-result}}

Six \dmeth\ lines have been detected toward Orion BN/KL, and three of them are slightly blended with other molecular lines. Four \dmeth\ lines are a-type ($\Delta K_a=0$ and $\Delta K_c=1$) R-branch ($\Delta J=1$) transitions with one b-type ($\Delta K_a=1$ and $\Delta K_c=1$) Q-branch ($\Delta J=0$) transition and one b-type R-branch transition. The \dmeth\ line parameter measurements toward Orion BN/KL are summarized in the appendix (Table \ref{table-CH2DOH}). 

\paragraph{\bf{Emission peaks:}} The \dmeth\ emission is mostly present in the southern part of the Orion BN/KL region, or the Compact Ridge region with a characteristic \vlsr\ of $\approx$ 8 \kms. For example, the \dmeth\ velocity channel maps in Figures \ref{Fig-CH2DOH-CH3OD-CH3OH-chmap} and \ref{Fig-CH2DOH-CH3OH-chmap} show one strong peak (deuterated methanol 1, dM-1) and two weaker peaks (dM-2 and dM-3) in the 7--9 \kms\ velocity range. 

There is no clear \dmeth\ detection at the Orion Hot Core itself (the strong dust continuum peak), except for the weak emission close to the source $I$ position \citep{Menten1995,Garay1987} shown in the \dmeth\ $5_{2,4}-4_{2,3}\ e_0$ line (see Fig. \ref{Fig-CH2DOH-CH3OD-CH3OH-chmap}, \vlsr=5--7 \kms). Since the feature associated with source $I$ is only seen in our strongest \dmeth\ line, it is not clear if the signal-to-noise ratio (S/N) is too low to detect this feature in other transitions, or if this feature is simply caused by contamination of other molecular lines from the region close to the Hot Core. Nevertheless, there is no indication of other molecular lines at this frequency. 

The \dmeth\ spectra toward the three main peaks are shown in the appendix (Figure \ref{Fig-CH2DOH-spectra}). The average \vlsr\ measured at the peak emission channel is $7.7\pm0.4$ \kms\ for dM-1, $7.8\pm0.4$ \kms\ for dM-2, and $7.8\pm0.4$ \kms\ for dM-3. The FWHM line widths estimated by fitting a Gaussian profile are $2.1\pm1.2$ \kms\ and $2.3\pm0.9$ \kms\ for dM-1 and dM-2, respectively. The weak dM-3 source exhibits a line width of approximately $3.7\pm0.5$ \kms, which is broader than the dM-1 and dM-2 line widths; this is likely because of the blending of two or more velocity components (see below).

Additionally, dM-1 has a counterpart in the methyl formate emission (MF-2) observed by \citet{Favre2011}, and dM-2 and dM-3 are nearly coinciding with MF-3 and MF-1, respectively. It should be noted that MF-1, the strongest methyl formate emission peak in Orion BN/KL \citep{Favre2011}, shows weak \dmeth\ emission in our data (dM-3). As shown by \citet{Favre2011}, MF-1 exhibits two velocity components at 7.5 \kms\ and 9.2 \kms, and these two components are only seen in the 225878.3 MHz \dmeth\ line toward dM-3 where our spectral resolution is higher (0.42 \kms). Furthermore, dM-2 and dM-3 are close to the \meth\ emission peak in the SMA \meth\ and \thmeth\ images ($\theta_{\rm syn}\approx1\farcs1\times0\farcs9$, $E_{\rm up}/k=110-730$ K) obtained by \citet{Beuther2005} around 337 GHz. It is interesting to note that dM-2 is close to the 25 GHz and 95 GHz \meth\ maser emission peaks reported by \citet{Matsakis1980} and \citet{Plambeck1988}, respectively.

\begin{table*}
\caption{Derived column density and temperature for \dmeth\ and \meth}             % title of Table
\label{table-summary}      % is used to refer this table in the text
\centering                          % used for centering table
\begin{tabular}{lccccccccc}        % centered columns (4 columns)
\hline\hline                 % inserts double horizontal lines
Position & R.A. & Dec. & Size    & $T_{\rm rot}$(\dmeth) & $N$(\dmeth) & $T_{\rm rot}$(\meth) &  $N$(\meth)  & [\dmeth]/[\meth]        \\    
         & $05^{\rm{h}}35^{\rm{m}}$ & $-05\degr22\arcmin$       &         & (K)                   & (\cmm)      & (K)                  & (\cmm)       &                   \\
       
\hline                        % inserts single horizontal line 

dM-1  & $...14\fs442$ & $...34\farcs86$ & $1\farcs8\times0\farcs8$&  130\tablefootmark{a} & $8.8\pm0.9\times10^{15}$ & ...                  & ...                      &  ...   \\
      & & & $3\farcs8\times2\farcs0$&  130\tablefootmark{a} & $4.7\pm0.5\times10^{15}$ & 130\tablefootmark{a} & $4.2\pm0.4\times10^{18}$ & $1.1\pm0.2\times10^{-3}$  \\
      & & & $3\farcs8\times2\farcs0$&   60                  & $2.4\pm0.3\times10^{15}$ & $60.3\pm3.7$         & $2.1\pm0.3\times10^{18}$ &  $1.1\pm0.2\times10^{-3}$  \\
dM-2  & $...14\fs320$& $...37\farcs23$ & $1\farcs8\times0\farcs8$&   85\tablefootmark{a} & $2.4\pm0.3\times10^{15}$ & ...                  & ...                      &  ...  \\
      & & & $3\farcs8\times2\farcs0$&   85\tablefootmark{a} & $1.7\pm0.1\times10^{15}$ & 85\tablefootmark{a}  & $2.2\pm0.7\times10^{18}$ &  $7.7\pm2.5\times10^{-4}$   \\
      & & & $3\farcs8\times2\farcs0$&   66                  & $1.4\pm0.2\times10^{15}$ & $66.2\pm4.7$         & $1.9\pm0.3\times10^{18}$ &  $7.4\pm1.6\times10^{-4}$   \\      
dM-3  & $...14\fs107$ & $...37\farcs43$ & $1\farcs8\times0\farcs8$&   58                  & $9.5\pm4.5\times10^{14}$ & ...                  & ...                      &  ...   \\
      & & & $3\farcs8\times2\farcs0$&   58                  & $1.5\pm0.1\times10^{15}$ & $57.6\pm1.8$         & $1.7\pm0.1\times10^{18}$ &  $8.8\pm0.8\times10^{-4}$  \\
KL-W  & $...14\fs159$ & $...28\farcs25$& $1\farcs8\times0\farcs8$&   44                  & $<3.0\times10^{14}$      & ...                  & ...                      &  ...   \\
      & & & $3\farcs8\times2\farcs0$&   44                  & $<2.0\times10^{14}$      & $43.7\pm4.3$         & $9.2\pm2.1\times10^{17}$ &  $<2.2\times10^{-4}$   \\
\hline                                   %inserts single line
\end{tabular}
\tablefoot{The [\dmeth]/[\meth] ratios are derived only for the $3\farcs8\times2\farcs0$ clump size. For dM-1 and dM-2, the \dmeth\ column densities are derived from the population diagrams, and \meth\ column densities and temperatures are derived for all sources from the population diagrams. The \dmeth\ column density of dM-3 is derived only from the $5_{2,4}-4_{2,3}$ $e_0$ line at 223422.3 MHz, assuming a rotational temperature of 58 K (estimated from \meth). The positions of dM-1, dM-2, dM-3, and KL-W are given at Epoch J2000.0.
\tablefoottext{a}{The methyl formate rotational temperatures derived by \citet{Favre2011} are adopted here for \dmeth\ and \meth.}
}

\end{table*}

\paragraph{\bf{Population diagrams:}} We use population diagrams to estimate the column density of \dmeth\ \citep[e.g.,][]{Goldsmith1999,Turner1991}

\begin{equation}
\ln\frac{N_{\rm up}}{g_{\rm up}}=\ln\frac{3k\int T_{\rm B}dV}{8\pi^3\nu S\mu^2}=\ln\frac{N}{Q}-\frac{E_{\rm up}}{kT_{\rm rot}},
\end{equation}
where $N$ is the total column density, $T_{\rm rot}$ the rotational temperature, $S$ the line strength, and $\mu$ the dipole moment \citep[$\mu_{x}=1.44$ D and $\mu_{z}=0.89$ D,][]{Ivash1953}. In this work, the partition function $Q$ is calculated according to our approximation $Q=2.25\ T^{1.5}$. We used the four \dmeth\ lines at 223 GHz ($e_0$ state) observed with the same spatial resolution in this population diagram analysis. Because our data points fall in a narrow energy range, the population diagrams give a temperature of about 100 K with a very large uncertainty. Hence, the rotational temperatures were assumed and taken from the work of \citet{Favre2011} to reduce the statistical error. We assumed here that both \meth\ and \mf\ trace the same gas because both lines show the same emission peaks, \vlsr, and $\Delta V$. Figure \ref{Fig-rotation-1} shows the least-squares fit result where we assumed that the rotational temperatures according to the \mf\ excitation temperatures are 130 K for dM-1 and 85 K for dM-2 as derived by \citet[][]{Favre2011} with the same spatial resolution ($1\farcs79\times0\farcs79$). The derived \dmeth\ column densities are given in Table 3. We find $8.8\pm0.9\times10^{15}$ \cmm\ and $2.4\pm0.3\times10^{15}$ \cmm\ for dM-1 and dM-2, respectively. Adopting an \HH\ column density of $3.1\times10^{24}$ \cmm\ as derived by \citet{Favre2011} from the 223 GHz dust continuum, we obtain a \dmeth\ relative abundance of $2.8\pm0.3\times10^{-9}$ toward dM-1. The relative \dmeth\ abundance is uncertain toward dM-2 because this clump does not have a strong dust continuum emission and no reliable \HH\ column density can be derived.

\citet{Jacq1993} derived a \dmeth\ column density of $2.6-5.4\times10^{15}$ \cmm\ toward Orion IRc2 with an estimated temperature of 88 K and an assumed source size of 15\arcsec\ from their IRAM 30m data. When smoothing our \dmeth\ data to a source size of 15\arcsec, the derived \dmeth\ column density is $0.2-1.4\times10^{15}$ \cmm, assuming a maximum missing flux of 50\% and an excitation temperature of 88 K toward IRc2. The column density disagrees with that of \citet{Jacq1993} by a factor of 2--3. Owing to the importance of deuteration ratios, which are obtained based mainly on previous single-dish estimates, we have revised the \dmeth\ column density calculations together with \methd\ in Appendix \ref{app1} to compare them with our interferometric results. We find that the \dmeth\ column density obtained by \citet{Jacq1993} is overestimated by a factor of 2--3, and the revisited value is thus consistent with our result mentioned above.

\subsection{\meth\label{meth-result}}

Three E-type methanol lines around 101 GHz (see Table 2) have been detected; the line parameters toward different peaks are summarized in Table \ref{table-CH3OH}. The \meth\ emission exhibits a V-shaped structure (Figs. \ref{Fig-CH2DOH-CH3OD-CH3OH-chmap} and \ref{Fig-CH2DOH-CH3OH-chmap}) similar to that observed for several other molecular lines and for dust continuum emission. The eastern side of the V-shaped structure follows the dense ridge, and the bottom part coincides with the Compact Ridge region. The western part of this V-shaped structure lies in the N-S direction, and one \meth\ emission peak (Orion KL western clump, KL-W, \citealp{Wright1992}; MF-4/5, \citealp{Favre2011}) is located to the north close to the IR source IRc6, which lies $\sim3\arcsec$ north of IRc7 \citep[see, e.g.,][]{Gezari1992,Gezari1998}. Because of the lower spatial resolution ($3\farcs79\times1\farcs99$) of these \meth\ lines, only dM-1, dM-3, and KL-W are spatially resolved. The average FWHM line width toward the \meth\ emission peaks is $4.4\pm1.2$ \kms, and the average line width at KL-W is about 6.5 \kms, which is likely due to the blending of two or more velocity components at 8 and 10--11 \kms\ close to MF-4 and MF-5 \citep{Favre2011}. In addition, the E-type methanol lines' average velocity is $8.2\pm0.9$ and $8.6\pm0.9$ \kms\ toward the three \dmeth\ emission peaks and KL-W, respectively. Since KL-W is located $\sim 7\arcsec$ from the Hot Core and $\sim 10\arcsec$ from the Compact Ridge, the 8 \kms\ velocity component observed in single-dish observations ($\theta_{\rm MB}>20\arcsec$) may contain a large amount of gas ($\sim25\%$ of our \meth\ total flux) from the KL-W region or the western side of the V-shaped structure. This will affect the analysis based only on the velocity components.

Because the \meth\ lines are optically thick in most cases \citep[e.g.,][]{Menten1988}, we first estimated the optical depth of the \meth\ lines detected here around 101 GHz. In the Rayleigh-Jeans approximation we derive the opacity from
\begin{equation}
\tau=-\ln\left[ 1-\frac{T_{\rm b}-T_{\rm bg}}{T_{\rm ex}-T_{\rm bg}} \right],
\end{equation}
where $T_{\rm b}$ and $T_{\rm ex}$ are the source brightness and excitation temperatures (using the derived rotational temperatures of 40--70 K), $T_{\rm bg}$ is the background emission temperature, including both the cosmic microwave background radiation (2.73 K) and the Orion dust emission (around 0.8--1.4 K from the PdBI data, correcting for a high interferometric filtering loss in continuum of 90\%). The average \meth\ optical depth is estimated to be $\lesssim0.2$ for the three 101 GHz \meth\ lines. Therefore, in the optically thin case, we can apply the population diagram method to estimate the E-type methanol column densities and rotational temperatures. The population diagrams are shown in Figure \ref{Fig-rotation-2} and the derived excitation temperatures and column densities are listed in Table \ref{table-summary}.

Rotational temperatures of 40--70 K and column densities of about $10^{18}$ cm$^{-2}$ are found toward selected clumps in Orion BN/KL, assuming an A/E methanol abundance ratio of 1.2 \citep{Menten1988}. The derived \meth\ column densities are higher than the IRAM 30m result of \citet{Menten1988}, who obtained a \meth\ column density of $3.4\pm0.3\times10^{17}$ \cmm\ with a rotational temperature of 128$\pm$10 K, including both narrow and broad components and assuming a source size of 25\arcsec. To compare ours with their results, we smoothed our \meth\ spectra to the beam size (25\arcsec) of the 30m telescope at 101--102 GHz (Fig. \ref{Fig-rotation-3}). We derived a \meth\ column density of $2.3\pm0.5\times10^{17}$ \cmm\ with a rotation temperature of about 60 K at the same position as used by \citet{Menten1988}. Additionally, by adopting a higher temperature of 130 K, we derived a \meth\ column density of $4.6\pm0.4\times10^{17}$ \cmm, which is close to the column density derived by \citet{Menten1988}.

Our methanol rotational temperature estimate is lower than the 100--150 K obtained by \citet{Menten1988} using the 30m telescope and lower than the 80 K of \citet{Neill2011} using CARMA, but lies at the low end of the 50--350 K obtained by \citet{Beuther2005} using SMA. Lower rotational temperatures may be due to the non-LTE behavior (or subthermal excitation, see $\S$ \ref{Herschel-compare}) of methanol. In addition, choosing only the lower-$E_{\rm up}$ transitions tends to result in a lower excitation temperature from population diagrams \citep[see, e.g.,][]{Blake1987,Parise2002}. Nevertheless, column densities are not very sensitive to rotational temperatures in this temperature range (Fig. \ref{Fig-temp-ntot}), and the differences in \meth\ column densities derived with different temperatures are less than a factor of about 3.

\subsection{\methd\ \label{methd-result}}

\begin{figure*}
\centering
\includegraphics[angle=-90,width=0.95\textwidth]{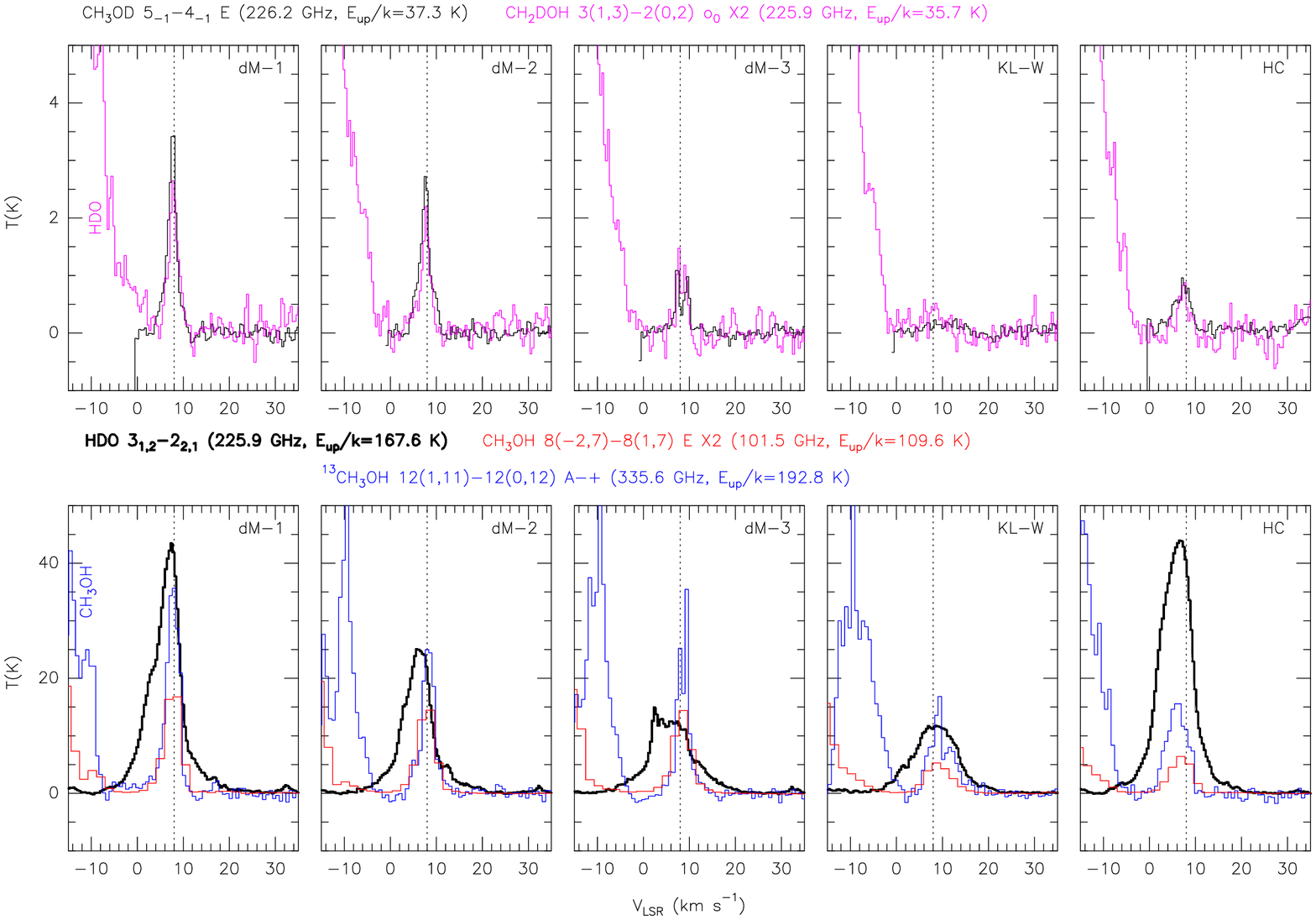}
\caption{Spectrum comparison of HDO, \meth, \thmeth, \dmeth, and \methd\ toward selected sources in Orion BN/KL with a similar resolution of about 3\arcsec. The SMA \thmeth\ line (Y.-W. Tang, priv. comm.) is plotted in blue for a synthesized beam of $3\farcs0\times2\farcs0$. The \meth\ line is plotted in red for a synthesized beam of $3\farcs8\times2\farcs0$. The HDO (thick black lines), \methd\ (thin black lines), and \dmeth\ (magenta) lines are plotted for a synthesized beam of $3\farcs6\times2\farcs3$. Dotted lines indicate a \vlsr\ of 8 \kms. Note that the spectrum intensities of \meth\ and \dmeth\ are multiplied by 2. Clearly, the spectra of methanol isotopologs have narrow line widths compared with those of HDO.}
\label{Fig-HDO-spectra}
\end{figure*}

A new transition of \methd\ ($5_{-1}-4_{-1}$ E) was detected at 226185.9 MHz for the first time in the ISM at the edge of our spectrometer and imaged toward Orion BN/KL. The \methd\ spectra for the selected sources in Orion BN/KL are shown in Figure \ref{Fig-HDO-spectra}, and the line parameters are listed in Table \ref{table-CH3OD}. The \methd\ channel maps are shown in Figure \ref{Fig-CH2DOH-CH3OD-CH3OH-chmap}. The \methd\ and \dmeth\ spatial distributions are very similar (see Fig. \ref{Fig-overlay-2} c), and both maps show the strongest emission at dM-1 with weaker emission toward dM-2 and dM-3. The line widths of \methd\ toward dM-1 and dM-2 are about 2--3 \kms, whereas two velocity components are clearly seen toward dM-3 with line widths of about 1--2 \kms.

\subsection{HDO \label{hdo-result}}

In one of our low angular resolution data sets ($3\farcs6\times2\farcs3$), we detected one HDO line at 225896.7 MHz ($3_{1,2}-2_{2,1}$). This is not sufficient for a detailed analysis, but it allows us to compare the spatial distribution and spectra between deuterated methanol and water emissions. Figure \ref{Fig-HDO-spectra} clearly shows that the optically thin lines from rarer isotopologs of methanol have similar spectral profiles with an average line width of about 3 \kms\ and LSR velocities of about 8 \kms\ at dM-1, dM-2, and dM-3. On the other hand, the HDO spectra in Orion BN/KL have broader line widths, which are likely composed by two or more velocity components. For example, the HDO spectra at dM-1 can be decomposed by two components at LSR velocities of about 5 and 7 \kms\ with line widths of about 10 and 4 \kms, respectively. The narrow line feature is likely produced by the same gas at which isotopologic methanol lines have been detected, and the broad line feature may be due mainly to the shock-heated gas associated with the Hot Core and source $I$ region where the hot ammonia and highly excited CH$_3$CN are present \citep[see, e.g.,][]{Goddi2011,Zapata2011}.

%________________________________________________________________
\section{Discussion}

\subsection{The \dmeth/\meth\ abundance ratios \label{meth-diss}}

           \begin{figure*}
   \centering
   \includegraphics[angle=-90,width=1\textwidth]{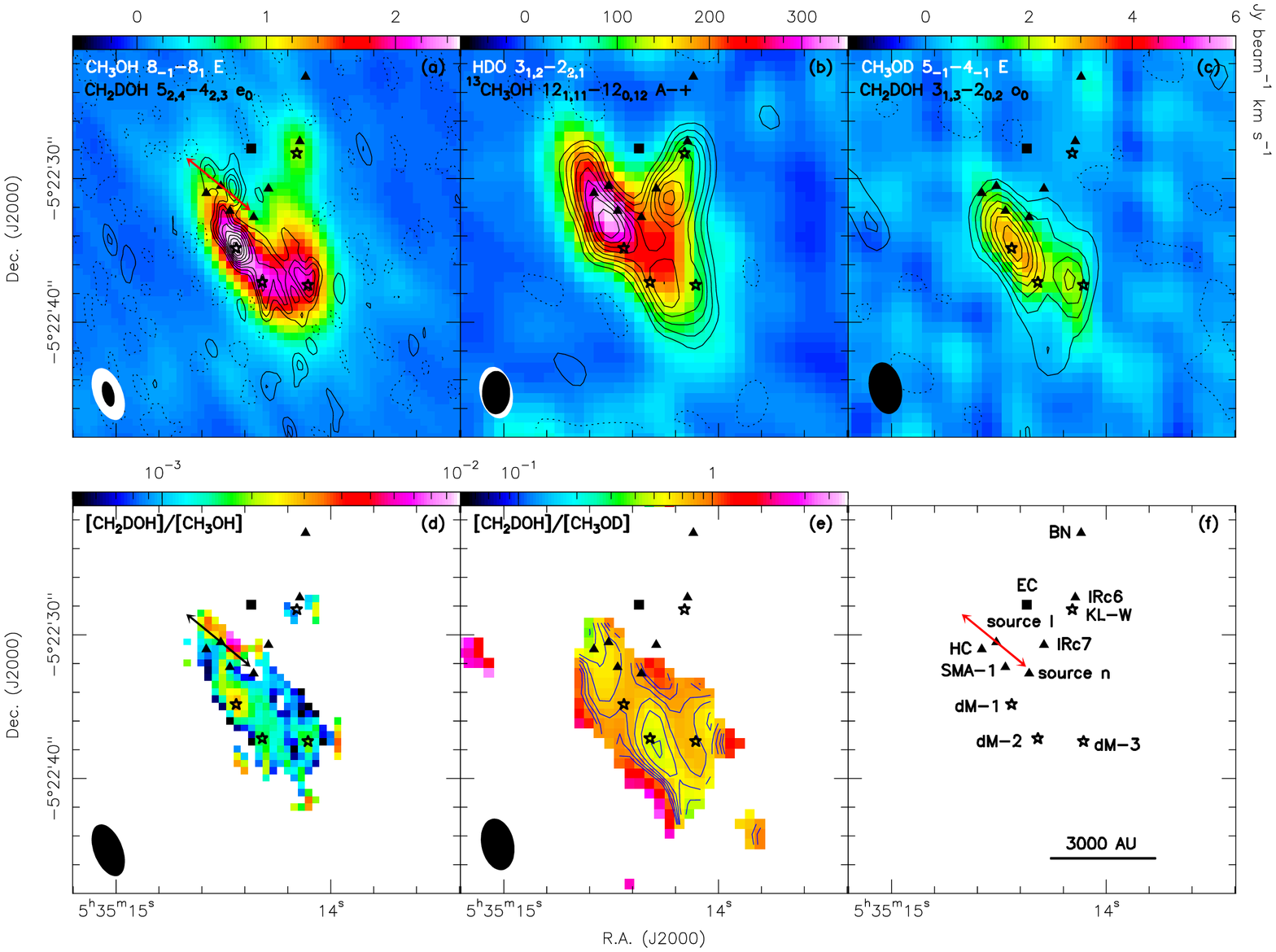}
   \caption{(a) Color image showing the integrated intensity map of the \meth\ $8_{-2}-8_{1}$ E line ($E_{\rm up}/k=109.6$ K, integrated from 5 to 11 \kms) for a synthesized beam of $3\farcs8\times2\farcs0$. Black contours represent the \dmeth\ $5_{2,4}-4_{2,3}$ line emission ($E_{\rm up}/k=48.4$ K, integrated from 5 to 11 \kms) and run from 15\% to 95\% in steps of 10\% of the peak intensity (1.0 \jb\ \kms) for a synthesized beam of $1\farcs8\times0\farcs8$. Black dashed contours correspond to --5\% of the peak intensity. (b) Color image showing the PdBI integrated intensity map of of the HDO $3_{1,2}-2_{2,1}$ line ($E_{\rm up}/k=167.6$ K, integrated from --8 to +17 \kms) for a synthesized beam of $3\farcs6\times2\farcs3$. Black contours represent the SMA \thmeth\ $12_{1,11}-12_{0,12}$ A--+ emission ($E_{\rm up}/k=192.8$ K, integrated from 0 to 12 \kms, Y.-W. Tang, priv. comm.), running from 20\% to 90\% in steps of 10\% of the peak intensity (316.6 \jb\ \kms) for a synthesized beam of $3\farcs0\times2\farcs0$. Black dashed contours correspond to --10\% of the peak intensity. (c) Integrated intensity map of the \methd\ $5_{-1}-4_{-1}$ E line  ($E_{\rm up}/k=37.3$ K, from 5 to 11 \kms) overlaid with the \dmeth\ $3_{1,2}-2_{0,2}\ o_{0}$ line ($E_{\rm up}/k=35.7$ K, integrated from 5 to 11 \kms) in black contours for a synthesized beam of $3\farcs6\times2\farcs3$. Black contours run from 0.28 (2 $\sigma$) to 1.12 \jb\ \kms\ in steps of 0.14 \jb\ \kms, and black dashed contours correspond to $-0.07$ \jb\ \kms.  (d) [\dmeth]/[\meth] abundance ratio map for a resolution of $3\farcs8\times2\farcs0$. The [\dmeth]/[\methd] abundance ratio is calculated by assuming the same excitation temperature (120 K) for both molecules. (e) [\dmeth]/[\methd] abundance ratio map for a resolution of $3\farcs6\times2\farcs3$. The [\dmeth]/[\methd] abundance ratio is calculated by assuming the same excitation temperature (120 K) for both \dmeth\ and \methd. Blue contours run from 0.4 to 1.0 in steps of 0.1. (f) The black square marks the center of explosion according to \citet{Zapata2009}. The positions of source BN, the Hot Core (HC), IRc6/7, and source $I$/n, and SMA-1 are marked by triangles. The positions of deuterated methanol emission peaks (dM-1, dM-2, and dM-3) and KL-W are marked by stars. The bipolar outflow of source $I$ is indicated.}
              \label{Fig-overlay-2}
    \end{figure*}

To investigate the \meth\ deuteration across the Orion BN/KL region, we smoothed our \dmeth\ data to the spatial resolution of the E methanol lines ($3\farcs8\times2\farcs0$) and derived the population diagram (Fig. \ref{Fig-rotation-1}) by using methanol rotational temperatures and \mf\ rotational temperatures for low and high temperature reference, respectively (see Table \ref{table-summary}, which summarizes our calculations toward different clumps in Orion BN-KL). The [\dmeth]/[\meth] abundance ratios are $1.1\pm0.2\times10^{-3}$, and $7.4\pm1.6\times10^{-4}$, and $8.8\pm0.8\times10^{-4}$ for dM-1, dM-2, and dM-3 with lower temperatures, respectively. An upper limit of about $2\times10^{-4}$ is given for the [\dmeth]/[\meth] abundance ratio at KL-W. It is interesting to note that the [\dmeth]/[\meth] ratio of dM-3 is higher than that of dM-2.

These abundance ratios do not depend much on the assumed $T_{\rm rot}$ (see Fig. \ref{Fig-temp-ntot}). If higher excitation temperatures are adopted (e.g., 130 K for dM-1 and 85 K for dM-2), the [\dmeth]/[\meth] ratios do not change much. These ratios (observed toward different clumps in Orion BN/KL, see Table \ref{table-summary}) are about one order of magnitude lower than the ratio (0.01--0.08) reported by \citet[][]{Jacq1993}. However, the [\dmeth]/[\meth] abundance ratios derived for the selected sources ($0.8-1.3\times10^3$) are consistent with a revised ratio of $0.8-2.8\times10^3$, where possible line blending and different observing positions were taken into account in the revisited calculations of the 30m observations (see Appendix \ref{app1} for more details). In addition, the [\dmeth]/[\meth] ratios in Orion BN/KL are lower than the deuterated methanol and formaldehyde abundance ratios estimated in several low-mass protostars, e.g., up to about 0.6 for [\dmeth]/[\meth] and 0.3 for [HDCO]/[H$_2$CO] in Class 0 sources \citep{Parise2006}. 

Assuming that both \dmeth\ and \meth\ are in LTE and have the same excitation temperature, we can produce a [\dmeth]/[\meth] ratio map by using two transitions of \dmeth\ and \meth\ and taking into account their line strengths, partition functions, and upper-state energies, the same method as used recently by \cite{Ratajczak2011}. In the [\dmeth]/[\meth] ratio map (Fig. \ref{Fig-overlay-2} d), dM-1 exhibits a higher ratio than observed toward the southern part of Orion BN/KL where dM-2 and dM-3 are located. It is interesting to note that the source $I$ region also shows higher \dmeth/\meth\ intensity ratios on each side of the bipolar outflow (marked with a double black arrow in the image). Since the feature associated with source $I$ is only seen in the strongest \dmeth\ line at 223422.3 MHz, more observations are needed to confirm this result. Because we derive an upper limit of $2.2\times10^{-4}$ for the [\dmeth]/[\meth] abundance ratio toward KL-W, which is well below the ratios derived for dM-1, dM-2, and dM-3, there might be different physical and/or chemical conditions at KL-W. Nevertheless, we cannot exclude yet that the non-detection of the \dmeth\ emission toward KL-W may result from spatial filtering in our PdBI interferometric data.

           \begin{figure*}
   \centering
   \includegraphics[angle=-90,width=0.9\textwidth]{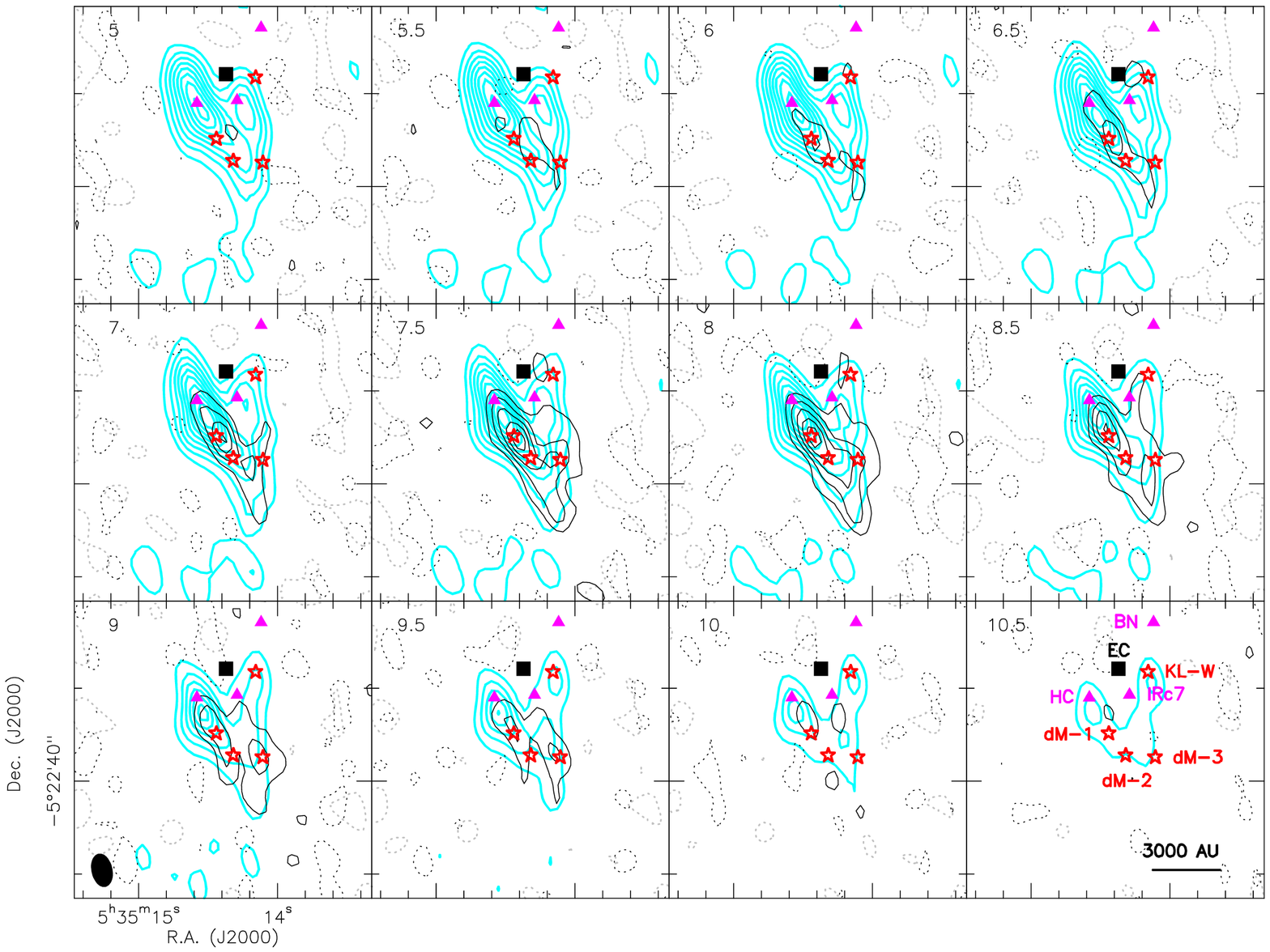}
   \caption{Channel maps of the \dmeth\ emission at 225878.3 MHz ($E_{\rm up}/k=35.7$ K) in black contours overlaid with the HDO $3_{1,2}-2_{2,1}$ line ($E_{\rm up}/k=167.6$ K) at 225896.7 MHz in light-blue contours for a synthesized beam of $3\farcs6\times2\farcs3$. The HDO emission contours run from $10\%$ to $90\%$ in steps of $10\%$ of the peak intensity (17.9 \jb), and the dashed contours represent --0.2 \jb\ (5 $\sigma$). The \dmeth\ emission contours run from 0.08 to 0.48 \jb\ in steps of 0.08 \jb, and the black dashed contours represent --0.04 \jb\ (1 $\sigma$). The black square marks the explosion center according to \citet{Zapata2009}. The positions of source BN, the Hot Core (HC), and IRc7 are marked by magenta triangles, and the positions of the deuterated methanol emission peaks (dM-1, dM-2, and dM-3) are marked by red stars.}
              \label{Fig-overlay-1}
    \end{figure*}

The deuteration ratios derived here are not markedly different from one clump to another except perhaps for KL-W (where an upper limit is derived). However, it is clear that the deuteration ratio tends to decrease from the dense clumps (dM-1 to dM-3) to the more extended gas in Orion BN/KL. The \HH\ column densities of dM-1 and dM-3 are about $3-5\times10^{24}$ \cmm, but the rotational temperature of \mf\ toward dM-1 (about 130 K) is higher than toward dM-2 and dM-3 (about 80 K). Therefore, the slight difference observed in the deuteration ratios is likely due to different gas temperatures of the clumps. Based on our present knowledge, we suggest that various competing processes could be at work. For instance, global, slow heating of the Orion BN/KL region through ongoing star formation should produce widespread effects and the expected net result is a fairly homogeneous deuterated ratio. On the other hand, heating by luminous infrared sources (e.g., in the vicinity of source $I$) could result in a higher temperature in this region, leading to a higher deuteration ratio toward dM-1.

\subsection{The \dmeth/\methd\ abundance ratios \label{methd-diss}}

Although only one \methd\ line ($5_{-1}-4_{-1}$ E at 226185.9 MHz, $E_{\rm up}/k=37.3$ K) has been detected in our observations, a similar [\dmeth]/[\methd] abundance ratio image can be produced in the same fashion as [\dmeth]/[\meth] by using the \dmeth\ $3_{1,3}-2_{0,2}\ o_0$ line ($E_{\rm up}/k=35.7$ K) at 225878.3 MHz. Figure \ref{Fig-overlay-2} (e) shows the [\dmeth]/[\methd] abundance ratio image by assuming the same excitation temperature for \dmeth\ and \methd, the method used for the [\dmeth]/[\meth] ratio map. This abundance ratio does not depend much on the temperature (see Fig. \ref{Fig-temp-ntot} b). Our results show that the [\dmeth]/[\methd] abundance ratios are less than unity in the central region of Orion BN/KL and the mean ratio is $0.7\pm0.3$, which also differs from the ratio (1.1--1.5) reported by \citet[][]{Jacq1993}. However, the revisited calculation as mentioned above (Appendix \ref{app1}) of the previous 30m observations of \dmeth\ \citep{Jacq1993}, \methd\ \citep{Mauersberger1988}, and \meth\ \citep{Menten1988} gives a lower [\dmeth]/[\methd] abundance ratio of $\lesssim0.6$ toward IRc2 (adopting a source size of 15\arcsec), which is consistent with our results and is much lower than the values derived in low-mass protostars \citep[e.g.,][]{Parise2002,Ratajczak2011}. Additionally, we note that the [\dmeth]/[\methd] abundance ratios are higher at dM-1 and dM-3 compared with that at dM-2.

The fact that the [\dmeth]/[\methd] ratios are less than unity in the central Orion BN/KL region is surprising because most of the theoretical and experimental studies on deuterated methanol predict more abundant \dmeth\ than \methd\ \citep[see, e.g.,][]{Charnley1997,Osamura2004,Nagaoka2005,Ratajczak2009}. Since \dmeth\ and \methd\ have very similar emission distributions in Orion BN/KL, their formations seem to be closely related. As shown by \citet{Osamura2004}, exchanging protons and deuterons on the two different parts of the methanol backbone is very inefficient in the post-evaporative gas-phase. However, the simulations of \citet{Osamura2004} show that the water vapor that evaporates from the surface above 80 K strongly affects the abundance of \methd, because the protonation of \water\ and HDO in gas phase leads to additional fractionation throughout the reaction network. Therefore, the observed abundance ratio between \dmeth\ and \methd\ might not reflect the primitive abundance ratio in the ice mantle of grain surfaces. This may explain the much lower [\dmeth]/[\methd] ratios seen in Orion BN/KL, where most of the regions have higher temperatures than 80 K. Besides, it indicates that \water\ and HDO may be two critical molecules for understanding the formation processes of deuterated methanol in Orion BN/KL.

%Additionally, the asymmetry on different dissociative recombination pathways for \dmeth\ and \methd\ has been discussed by \citet{Osamura2004} that the gas-phase \methd\ can be removed via
%\begin{equation}  
%{\rm CH_3OHD^++e^- \rightarrow CH_3OH+D, CH_3OD+H},
%\end{equation}
%while \dmeth\ cannot be destroyed in the same way due largely to the inefficiency of D-H exchange between the C- and O-ends of methanol. Furthermore, some studies on dissociative recombination of HDO$^+$ \citep{Jensen1999} and the dissociation of HDO \citep{Sayler2006} by fast ions showed the preference of breaking the O-H bond over the O-D bond. Therefore, the ${\rm CH_3OD+H}$ channel may be more efficient than the ${\rm CH_3OH+D}$ channel in the similar way.
%}

\subsection{CH$_3$OH subthermal excitation?  \label{Herschel-compare}}

Recently, \citet{Wang2011} detected more than 300 methanol lines around 524 GHz ($\theta_{\rm MB}\approx43\arcsec$) and 1061 GHz ($\theta_{\rm MB}\approx20\arcsec$) with the {\it Herschel}/HIFI instrument. Although these data were obtained with a broad beamwidth, it may be useful to investigate the implications of the Wang et al. results on our own data in terms of \meth\ excitation. The narrow and broad \meth\ components observed by Wang et al. show average FWHM line widths of 2.7 and 4--11 \kms\ at $V_{\rm LSR}=8.6$ and 6--9 \kms, respectively. Their LSR velocities are consistent with our data. However, their broad components' line widths are wider than ours. This is likely due to optical depth broadening ($\tau\sim15$ in the {\it Herschel}/HIFI \meth\ data), although we cannot exclude that some faint broad and extended velocity components detected by {\it Herschel} may also be partly filtered out in our interferometric data.

Our three \meth\ 101 GHz transitions include upper energy levels with $E_{\rm up}/k\lesssim110$ K, from which we derived $T_{\rm rot}=40-70$ K. The \meth\ lines in this energy level range detected by {\it Herschel} \citep[see, e.g., Figure 4 in][]{Wang2011} also give a similar trend: a low rotational temperature and a high column density are derived by using only the low $E_{\rm up}$ data in the \meth\ population diagram. The \meth\ lines from the {\it Herschel}/HIFI observations are optically thick, but the $^{13}$\meth\ lines ($\tau=0.03-0.28$) also show the same trend. This suggests that our low-temperature result ($T_{\rm rot}=40-70$ K) is unlikely to be caused by optical depth effects. Because the LSR velocities of our \meth\ lines are consistent with the {\it Herschel}/HIFI data, it is more likely that our low \meth\ excitation temperatures are not caused by a cold component along the line of sight, but that they are largely generated by subthermal excitation conditions.

\subsection{Comparison with the HDO emission\label{HDO-compare}}

In Figure \ref{Fig-overlay-1}, the superposition of HDO $3_{1,2}-2_{2,1}$ and \dmeth\ $3_{1,3}-2_{0,2}$ lines shows a globally similar distribution for different velocity channels. This result is significant because both lines were detected in the same PDBI data set around 225.9 GHz. However, details differ, as one may expect for two transitions with different upper energy levels (168 K and 36 K for HDO and \dmeth, respectively), and for different molecular formation scenarios, which imply, for instance, that methanol is formed on grain surfaces at higher densities than water.

The HDO emission peaks between the Orion Hot Core and dM-1, near the dust clump SMA-1 \citep[][; see also our Fig. \ref{Fig-overlay-2}]{Beuther2004}. Dust emission toward this peak has been detected with SMA in the 870 $\mu$m band by \citet{Tang2010}. Owing to a position offset of about 1\arcsec, it is not clear whether this clump coincides with the SMA-1 position observed by \citet{Beuther2004} with a resolution of $0\farcs78\times0\farcs65$. In addition to the prominent emission peak near SMA-1, the HDO emission shows a similar V-shaped distribution as \meth. There is HDO emission close to the IR source IRc7, but it is interesting to note that the HDO emission does not peak at IRc7. It peaks to the southwest of IRc7 instead. A similar situation is observed with SMA for the \thmeth\ emission (see the right panel of Figure \ref{Fig-overlay-2}; Y.-W. Tang, priv. comm.). In addition, it seems that the overall \dmeth\ emission distribution is shifted to the south by a few arcseconds with respect to the HDO emission. For example, the strongest HDO emission peak, which corresponds to the NH$_3$ column density peak derived by \citet{Goddi2011}, is located 3\arcsec\ north of dM-1. This shift is clearly seen in Figure \ref{Fig-overlay-2} where the HDO and \thmeth\ emissions lie farther north than the \meth\ and \dmeth\ emissions. The different spatial distributions observed for \dmeth\ and \meth\ on one side, and HDO and \thmeth\ on the other side (see Figs. \ref{Fig-overlay-2} and \ref{Fig-overlay-1}) cannot be attributed to optical depth effects because (a) we have shown that our \dmeth\ (1.3 mm emission) and \meth\ (3 mm emission) data are optically thin; (b) our estimate of the HDO line opacity is $\lesssim0.3$ \citep[agrees with][]{Jacq1990}. The difference observed in the distribution of these molecular species could be due to different temperatures across the Orion molecular material involving processes such as heating by luminous infrared sources or shocks.

\subsection{Deuteration ratios of water and methanol\label{deuteration-ratio}}

The [HDO]/[H$_2$O] abundance ratio has been reported to be about a few times $10^{-4}$ toward Orion BN/KL \citep[see, e.g.,][]{Pardo2001,Jacq1990}, which is about one order of magnitude lower than the [\dmeth]/[\meth] ratio derived here. [HDO]/[H$_2$O] ratios lower than the [\dmeth]/[\meth] ratios have also been reported toward low-mass protostars \citep{Parise2005,Liu2011}. Since heavy water (D$_2$O) is much less abundant than HDO, i.e., [D$_2$O]/[HDO]=$1.7\times10^{-3}$ in IRAS 16293--2422 \citep{Butner2007}, HDO seems to be the main deuterium  (gas phase) reservoir for water in Orion BN/KL. For methanol, \methd\ is as abundant as \dmeth, and other doubly or multiply deuterated methanol species are expected to be less abundant. Therefore, the total deuteration ratio for methanol (including \dmeth\ and \methd) in Orion BN/KL is about one order of magnitude higher than that of water (including HDO and D$_2$O). If both water and methanol sublimate from ice mantles with similar time scales, and supposing that their destruction rates are comparable, the higher abundance of deuterated methanol may just reflect different chemical reaction speeds on dust surface. At a later stage, formation and destruction rates in the gas phase will of course complicate the analysis. Some experiments have been conducted to investigate this question. For example, H-D exchange between water and methanol has been proposed by several authors in different physical conditions, e.g., temperature increase and ultraviolet light irradiation \citep{Ratajczak2009,Weber2009}. In addition, the enrichment of deuterated methanol has been experimentally reproduced by H-D substitution in solid methanol at 10 K. However, a high atomic D/H ratio of 0.1 was required and the substitution was not seen in water and ammonia \citep[][]{Nagaoka2005}. Therefore, H-D substitution may qualitatively explain why methanol is easier to be deuterated than water; but more chemical experiments and modeling are clearly needed to quantitatively address this question.

%______________________________________________________________

\section{Conclusions}

The main findings and conclusions of our study based on observations of several transitions of deuterated methanol and one transition of deuterated water in Orion BN/KL are as follows.

\begin{enumerate}
 \item We have obtained the first high angular resolution ($1\farcs8\times0\farcs8$) \dmeth\ images detected around 223.5 GHz toward Orion BN/KL and compared these data with somewhat lower resolution ($3\farcs8\times2\farcs0$) \meth\ images at 101.5 GHz. The strongest \dmeth\ and \meth\ emissions come from the Hot Core southwest region exhibiting an LSR velocity of about 8 \kms, typical of the Orion Compact Ridge region. The \dmeth\ emission is clumpy and the column densities are estimated to be about $1-9\times10^{15}$ \cmm\ toward these clumps. The \meth\ column densities are about $3-5\times10^{17}$ \cmm\ across Orion BN/KL, leading to a [\dmeth]/[\meth] deuteration ratio of $0.8-1.3\times10^{-3}$ toward three deuterated methanol clumps and below $2\times10^{-4}$ toward KL-W.
 \item The [\dmeth]/[\methd] abundance ratio map was obtained for Orion BN/KL, and their ratios are less than unity at the central part of the region. These ratios are lower than the statistical factor of 3 derived in the simplest deuteration models, and definitely lower than the values derived in low-mass protostars \citep[e.g.,][]{Parise2002,Ratajczak2011}.
 \item We have mapped with moderately high spatial resolution ($3\farcs6\times2\farcs3$) the 225.9 GHz transition of HDO and compared its distribution with \dmeth, \meth, and \thmeth. We find that the deuterated water ratio is about one order of magnitude lower than the deuterated methanol ratio. H-D substitution may explain why methanol is easier to be deuterated than water.   
  \item The deuteration ratios derived in this work are not strongly different from one clump to another, except perhaps toward KL-W where more observations are desirable to conclude. However, to explain the slight differences observed locally in the abundance ratios of identified clumps, we suggest that various processes could be competing, for instance, heating by luminous infrared sources, or heating by shocks. 
\end{enumerate}

%______________________________________________________________

\begin{acknowledgements}
We thank Y.-W. Tang for kindly providing the SMA \thmeth\ image. T.-C. Peng acknowledges support from the ALMA grant 2009-18 at Universit\'{e} de Bordeaux1/LAB. B. Parise is supported by the German Deutsche Forschungsgemeinschaft, DFG Emmy Noether project number PA1692/1-1. We thank the referee, who helped us in restructuring our initial work and in clarifying our main conclusions.
\end{acknowledgements}

%
%---------------------------------------------------------------------------
%

\Online

\begin{appendix}

\section{Revisiting the \dmeth/\methd\ abundance ratios in Orion BN/KL \label{app1}}

To compare our high angular resolution data with previous single-dish observations, we re-calculated the temperatures and column densities obtained by \citet{Mauersberger1988} and \citet{Jacq1993} from their 30m data for \methd\ and \dmeth, respectively. The line parameters of \methd\ and \dmeth\ used in the calculations are listed in Table \ref{table-methd-dmeth}, and the population diagrams are shown in Figure \ref{Fig-CH2DOH-CH3OD-fit}.

\subsection*{\methd}

\begin{figure}
\centering
\includegraphics[angle=-90,width=0.48\textwidth]{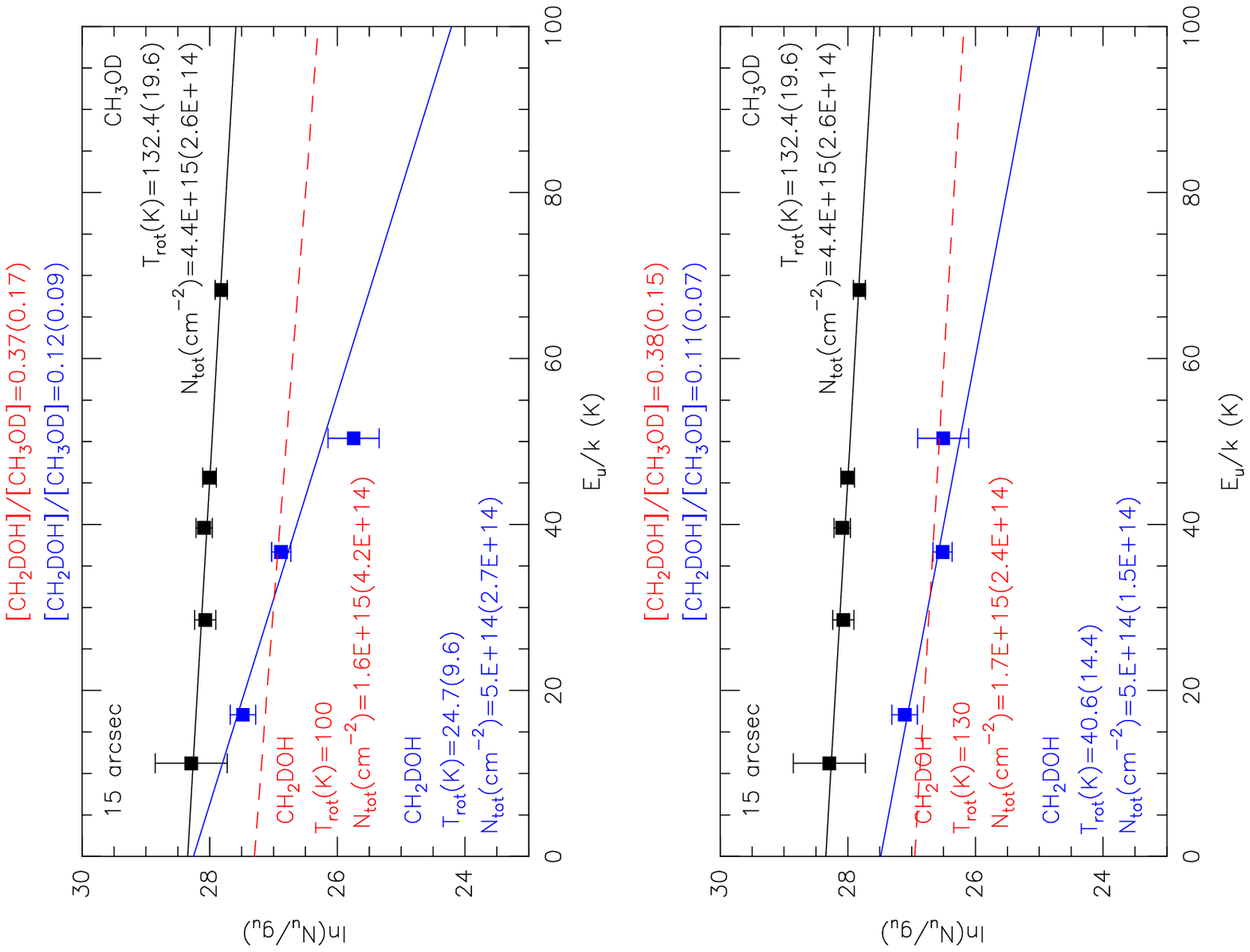}
\caption{Population diagrams of \methd\ and \dmeth\ using the 30m data of \citet{Mauersberger1988} and \citet{Jacq1993}, respectively. The old \dmeth\ effective line strengths of \citet{Jacq1993} (upper panel) and the new ones (lower panel) were used in the calculations. The beam filling factor was taken into account by assuming a source size of 15\arcsec. The rotational temperatures and column densities were estimated by a least-squares fit, and the results and uncertainties are given in the diagrams. For \dmeth, the red dashed line in the upper panel indicates the fit with a fixed temperature of 100 K as adopted by \citet{Jacq1993} and the one in the lower panel the fit with a temperature of 130 K (the same as the \methd\ temperature derived here).}
\label{Fig-CH2DOH-CH3OD-fit}
\end{figure}

We selected only five clean \methd\ lines (without any apparent line-blending) around 140--156 GHz detected by \citet{Mauersberger1988} with a similar resolution of 16\arcsec--17\arcsec\ to reduce the fitting uncertainty. The \methd\ lines detected by \citet{Jacq1993} were excluded because no \methd\ spectra can be used to judge the possible line-blending and data quality. Effective line strengths $S\mu^2$ of the \methd\ lines were taken from \citet{Anderson1988}, and the partition function $Q=1.41\ T^{1.5}$ was used. The rotational temperature (132.4$\pm19.6$ K) and column density ($4.4\pm0.3\times10^{15}$ \cmm) were estimated by a least-squares fit (Fig. \ref{Fig-CH2DOH-CH3OD-fit}), assuming a source size of 15\arcsec. The revised \methd\ temperature and column density are consistent with those of \citet{Mauersberger1988}, who derived a rotational temperature of 50--150 K and a column density of $1-5\times10^{15}$ \cmm.

\subsection*{\dmeth}

For \dmeth, we judge that only the three detections reported by \citet{Jacq1993} can be reliably used in the population diagrams (see Table \ref{table-methd-dmeth}). Both old \dmeth\ $S\mu^2$ values taken from \citet{Jacq1993} and new ones (see Table \ref{table-methd-dmeth}) provided by B. Parise (priv. comm.) were used in the calculations. (Fig. \ref{Fig-CH2DOH-CH3OD-fit}). The \dmeth\ partition function $Q=2.25\ T^{1.5}$ and a source size of 15\arcsec\ were used here.

The new values of \dmeth\ $S\mu^2$ employed in this work are based on those tabulated in the paper by \citet{Parise2002} but a factor of three higher. This revision corrects for an inconsistency with the partition function reported by \citet{Parise2002}, and some dipole moment values can be quite uncertain, which may explain the discrepancy with the values used by \citet{Jacq1993} and why they were never officially released in the JPL database. However, as shown below, both the old $S\mu^2$ values used by \citet{Jacq1993} and new ones adopted here give similar estimates of the \dmeth\ column densities in Orion BN/KL. We believe that the uncertainties of the \dmeth\ $S\mu^2$ values of our present work are not dominant in the column density calculations. Nevertheless, a new spectroscopic study of \dmeth\ with a robust fitting of the Hamiltonian would definitely give better confidence in these intensities.

Using the old $S\mu^2$, a least-squares fit to the \dmeth\ data shows a low rotational temperature (about 25$\pm$10 K) with a column density of $5.0\times10^{14}$ \cmm, consistent with the first estimate of \citet{Jacq1993}, who derived a rotational temperature of 34$^{+60}_{-13}$ K and a column density of $6\times10^{14}$ \cmm. However, since the rotational temperatures derived by \citet{Menten1988} and \citet{Mauersberger1988} for \meth\ and \methd\ are about 100 K toward Orion BN/KL, \citet{Jacq1993} also adopted a higher temperature for \dmeth. They obtained a \dmeth\ column density of $3.9\times10^{15}$ \cmm, which is 2.6 times higher than our revised value ($1.6\times10^{15}$ \cmm, see Fig. \ref{Fig-CH2DOH-CH3OD-fit} upper panel). This is because they appear to correct the beam-filling factor twice in their $N_{l}/g_{l}$ calculations and population diagram. Therefore, the \dmeth\ column density derived by \citet{Jacq1993} with a temperature of 100 K was overestimated by a factor of 2--3.

With the new \dmeth\ $S\mu^2$, we derive a somewhat higher rotational temperature ($40.6\pm14.4$ K) and a column density of $5.0\times10^{14}$ \cmm\ similar to that derived by using the old $S\mu^2$. This suggests that the uncertainties in the \dmeth\ effective line strengths do not strongly affect the \dmeth\ column density estimate in this case.

\subsection*{\dmeth/\methd\ abundance ratios}

The overestimated \dmeth\ column density derived by \citet{Jacq1993} leads to a higher [\dmeth]/[\methd] abundance ratio (1.1--1.5) toward IRc2. The new calculation indicates that the [\dmeth]/[\methd] abundance ratio is 0.2--0.5 with a source size of 15\arcsec, assuming \dmeth\ and \methd\ have the same temperature of about 130 K. For a lower \dmeth\ temperature (i.e., about 40 K), the [\dmeth]/[\methd] ratio is even lower (about 0.1).

\begin{table*}
\caption{30m \methd\ and \dmeth\ line parameters }             % title of Table
\label{table-methd-dmeth}      % is used to refer this table in the text
\centering                          % used for centering table
\begin{tabular}{llrccccc}        % centered columns (4 columns)
\hline\hline                 % inserts double horizontal lines
Frequency & Transition             & $E_{\rm up}/k$ & $S\mu^2$  &  $\int T_{\rm MB}dV$  & $f$\tablefootmark{c} & $N_{\rm u}/g_{\rm u}$\tablefootmark{d}  & Comment\\    
(MHz)     & ($J_{k_{a},k_{c}}$)    & (K)            &  (D$^2$)  & (K km s$^{-1}$)       &                      &  (cm$^{-2}$)                            & \\

\hline
\multicolumn{8}{c}{CH$_3$OD \tablefootmark{a}} \\       
\hline                        % inserts single horizontal line

140175.20 & $4_{1,3}-4_{0,4}$ A$--$                   & 19.8 &  9.3 &  $5.1\pm0.7$  & 0.42 & $1.55\pm0.25\times10^{12}$ & \\  
143741.65 & $5_{1,4}-5_{0,5}$ A$--$                   & 27.5 & 11.2 &  $6.6\pm0.7$  & 0.43 & $1.57\pm0.20\times10^{12}$ &  \\     
148359.77 & $6_{0,6}-5_{1,5}$ A$++$                   & 31.7 &  5.7 &  $3.3\pm0.3$  & 0.45 & $1.45\pm0.16\times10^{12}$ &  \\
153324.00 & $7_{1,6}-7_{0,7}$ A$--$                   & 47.4 & 14.7 &  $7.6\pm0.6$  & 0.47 & $1.21\pm0.11\times10^{12}$ &  \\
155533.08 & $1_{1,0}-0_{0,0}$ E                       &  7.8 &  1.0 &  $0.9\pm0.4$  & 0.47 & $1.93\pm1.09\times10^{12}$ &  \\

\hline
\multicolumn{8}{c}{CH$_2$DOH\tablefootmark{b}} \\
\hline

99672.23  & $6_{1,5}-6_{0,6}$ $e_{\rm 0}$             & 50.4 & 12.3\tablefootmark{e} &  $0.3\pm0.1$  & 0.27 & $1.51\pm0.61\times10^{11}$&  \\     
136151.26 & $3_{1,2}-2_{1,1}$ $e_{\rm 0}$             & 17.1 &  2.1\tablefootmark{e} &  $0.6\pm0.1$  & 0.41 & $8.58\pm1.72\times10^{11}$& \tablefootmark{g} \\   
226818.36 & $5_{1,4}-4_{1,3}$ $e_{\rm 0}$             & 36.7 &  3.8\tablefootmark{e} &  $1.6\pm0.2$  & 0.66 & $4.72\pm0.71\times10^{11}$& \tablefootmark{g} \\

\hline

99672.23  & $6_{1,5}-6_{0,6}$ $e_{\rm 0}$             & 50.4 & 5.8\tablefootmark{f} &  $0.3\pm0.1$  & 0.27 & $3.23\pm1.29\times10^{11}$ &  \\     
136151.26 & $3_{1,2}-2_{1,1}$ $e_{\rm 0}$             & 17.1 & 3.1\tablefootmark{f} &  $0.6\pm0.1$  & 0.41 & $5.91\pm1.18\times10^{11}$ & \tablefootmark{g} \\   
226818.36 & $5_{1,4}-4_{1,3}$ $e_{\rm 0}$             & 36.7 & 5.5\tablefootmark{f} &  $1.6\pm0.2$  & 0.66 & $3.27\pm0.49\times10^{11}$ & \tablefootmark{g} \\

\hline                                   %inserts single line
\end{tabular}
\tablefoot{
\tablefoottext{a}{\citet{Mauersberger1988}; centered at KL-W$_{\rm M}$ ($05^{\rm{h}}35^{\rm{m}}14\fs18$, $-05\degr22\arcmin26\farcs5$, J2000).}
\tablefoottext{b}{\citet{Jacq1993}; centered at IRc2 ($05^{\rm{h}}35^{\rm{m}}14\fs47$, $-05\degr22\arcmin30\farcs2$, J2000).}
\tablefoottext{c}{The beam filling factor $f=\theta_{\rm s}^2/(\theta_{\rm s}^2+\theta_{\rm b}^2)$ is calculated by assuming a source size $\theta_{\rm s}$ of 15\arcsec. $\theta_{\rm b}$ is the 30m beam size.}
\tablefoottext{d}{The beam filling factor is taken into account in the calculation of $N_{\rm u}$, and the rms noise and calibration uncertainties are included (assuming $20\%$).}
\tablefoottext{e}{The Old $S\mu^2$ taken from \citet{Jacq1993}}
\tablefoottext{f}{The New $S\mu^2$ provided by B. Parise (priv. comm.).}
\tablefoottext{g}{The data obtained at the offset position (0\arcsec,$-$6\arcsec) of IRc2.}
}

\end{table*}

In addition, the \methd\ data of \citet{Mauersberger1988} were taken at the position close to KL-W \citep[KL-W$_{\rm M}$ hereafter,][]{Menten1988} instead of IRc2, which was used as the central position by \citet{Jacq1993}. According to our \methd\ image (Fig. \ref{Fig-overlay-2}), the intensity observed toward IRc2 is about twice as high as that of KL-W$_{\rm M}$, assuming a 30m telescope beam size of about 18\arcsec\ at 140 GHz and the same filtering. Therefore, the \methd\ abundance in Orion BN/KL derived by \citet{Mauersberger1988} may be underestimated by 50\% at most. A similar calculation (Fig. \ref{Fig-rotation-3}) also shows that the \meth\ column density toward KL-W$_{\rm M}$ is about 20\% lower than that toward IRc2.

In short, we conclude that the revisited [\dmeth]/[\methd] abundance ratio derived from the previous 30m observations is $\lesssim0.6$, and the [\dmeth]/[\meth] abundance ratio is estimated  to be $0.8-2.8\times10^{-3}$ toward IRc2, by adopting a \meth\ column density of $4.7\pm0.3\times10^{17}$ \cmm\ \citep{Menten1988} and correcting the possible density underestimate of about 20\% toward IRc2 (for a source size of 15\arcsec).

\end{appendix}

%\pagebreak
%\clearpage

%
%---------------------------------------------------------------------------
%

\begin{appendix} %First online appendix
\section{Complementary figures and tables \label{app2}}

   \begin{figure*}
   \centering
   \includegraphics[angle=-90,width=0.75\textwidth]{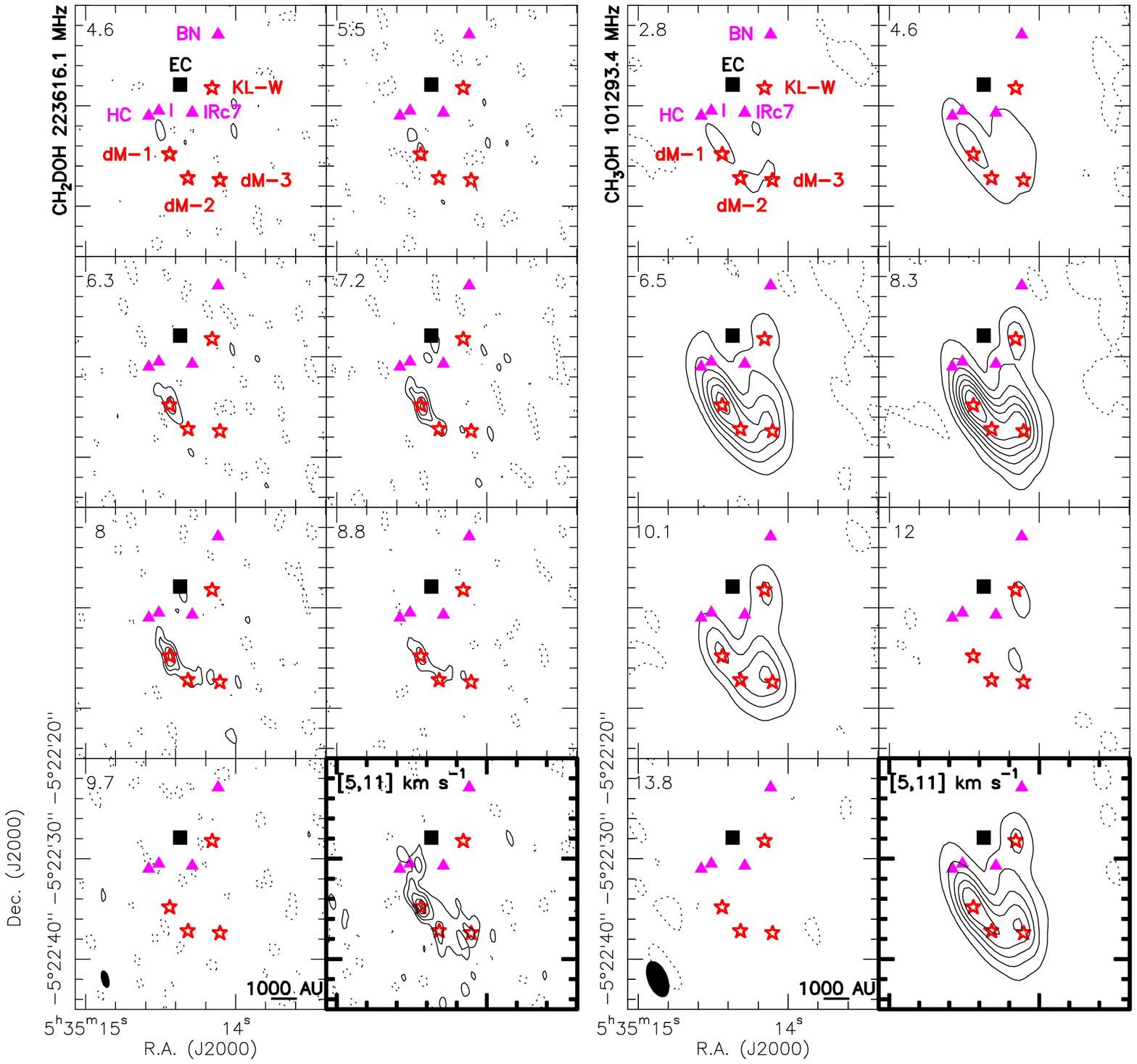}
   \caption{Left: Channel maps of the \dmeth\ doublet ($E_{\rm up}/k=48.4$ K) emission at 223616.1 MHz for a synthesized beam of $1\farcs8\times0\farcs8$. Contours run from 40 \mjb\ (2 $\sigma$) to 160 \mjb\ (2.77 K) in steps of 40 \mjb, and the dashed contours represent --20 \mjb. The bottom-right panel shows the integrated intensity (from 5 to 11 \kms) in contours running from 0.15 to 0.45 \jb\ \kms\ in steps of 0.15 \jb\ \kms, and the dashed contours represent --0.08 \jb\ \kms. Right: Channel maps of the \meth\ $7_{-2}-7_{1}$ E ($E_{\rm up}/k=91.0$ K) emission at 101293.4 MHz for a synthesized beam of $3\farcs8\times2\farcs0$. Contours run from 50 to 350 \mjb\ in steps of 50 \mjb. The dashed contours represent --30 \mjb (1 $\sigma$). The bottom-right panel shows the integrated intensity of \meth\ (from 5 to 11 \kms) in contours running from 0.3 to 1.5 \jb\ \kms\ in steps of 0.3 \jb\ \kms, and the dashed contours represent --0.03 \jb\ \kms. The black square marks the center of explosion according to \citet{Zapata2009}. The positions of source BN, the Hot Core (HC), IRc7, and source $I$ are marked by triangles. The positions of deuterated methanol emission peaks (dM-1, dM-2, and dM-3) and KL-W are marked by red stars.}
              \label{Fig-CH2DOH-CH3OH-chmap}
   \end{figure*}

\begin{figure*}
\centering
\includegraphics[angle=-90,width=0.9\textwidth]{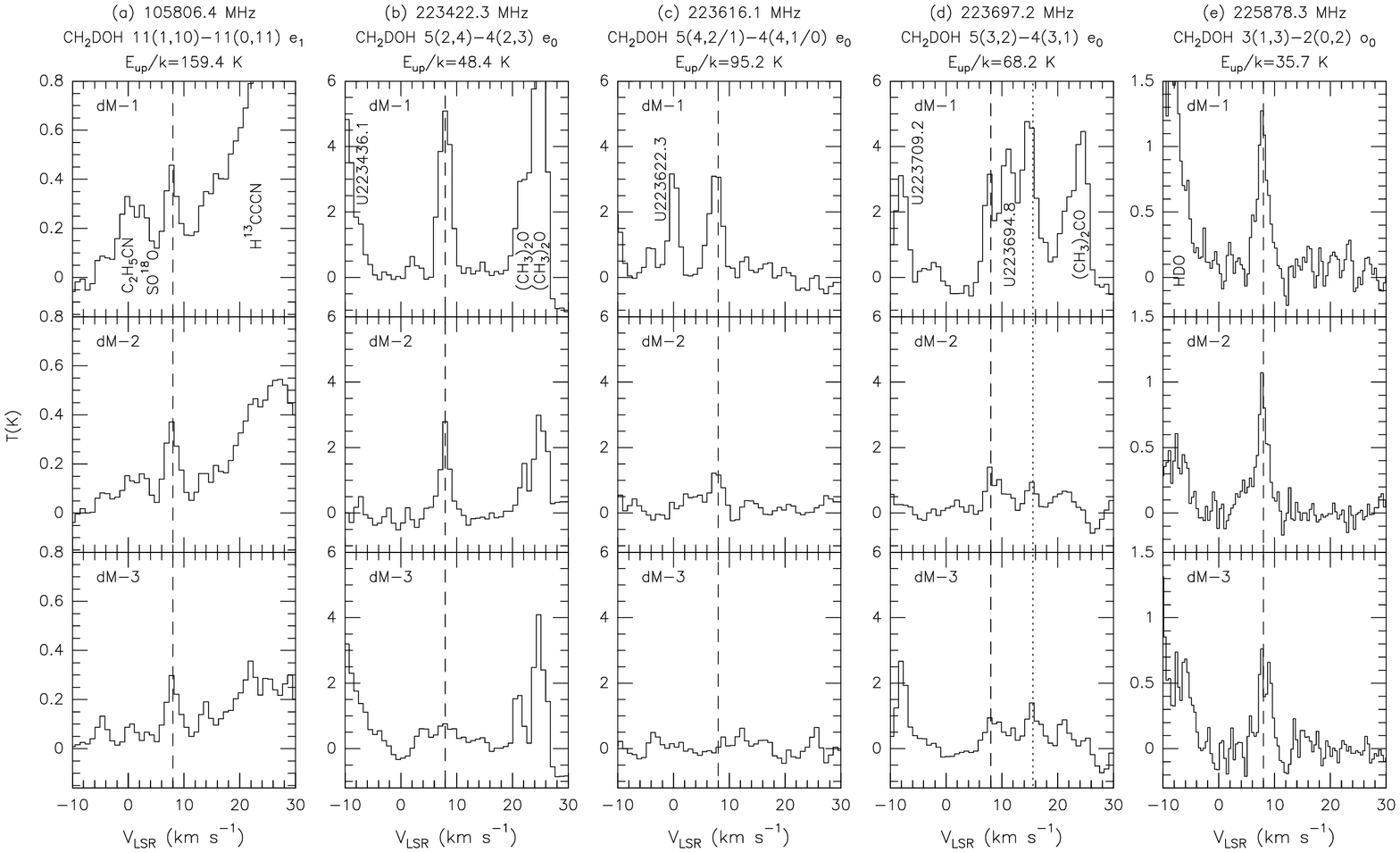}
\caption{\dmeth\ spectra toward dM-1, dM-2, and dM-3 for a synthesized beam of $1\farcs8\times0\farcs8$. The dashed lines indicate a \vlsr\ of 8 \kms. The unidentified molecular lines are denoted with their frequencies. (a) For dM-1, the \dmeth\ $11_{1,10}-11_{0,11}\ e_{1}$ line is blended with a broad H$^{13}$CCCN $J=12-11$ line ($\Delta V\approx10$ \kms) at 105799.1 MHz. The rms noise (1 $\sigma$) is 0.02 K. (b)-(d) The dotted line indicates the \dmeth\ $5_{3,3}-4_{3,2}$ $e_0$ line at 223691.5 MHz, which is blended with the (CH$_3$)$_2$CO $17_{7,11}-16_{6,10}$ EA/AE lines at 223692.1 MHz. The rms noise (1 $\sigma$) is 0.35 K. (e) The strong line to the left is the HDO $3_{1,2}-2_{2,1}$ line at 225896.7 MHz. Two velocity components at 7.8 and 9.0 \kms\ are seen toward dM-3. The rms noise (1 $\sigma$) is 0.12 K.}
\label{Fig-CH2DOH-spectra}
\end{figure*}

%\pagebreak

\begin{table*}
\caption{PdBI \dmeth\ line parameters at dM-1, dM-2, and dM-3}             % title of Table
\label{table-CH2DOH}      % is used to refer this table in the text
\centering                          % used for centering table
\begin{tabular}{lcccccccc}        % centered columns (4 columns)
\hline\hline                 % inserts double horizontal lines
Frequency\tablefootmark{a}  & Transition             & $E_{\rm up}/k$ & $S\mu^2$  & $T_{\rm peak}$  & $V_{\rm LSR}$  & $\Delta V$    & $\int TdV$  & Comment\\    
(MHz)      & ($J_{k_{a},k_{c}}$)    & (K)          &  (D$^2$)  &     (K)         & (km s$^{-1}$)  & (km s$^{-1}$) & (K km s$^{-1}$)     & \\

\hline
\multicolumn{9}{c}{dM-1} \\       
\hline                        % inserts single horizontal line

105806.4100 & $11_{1,10}-11_{0,11}$ $e_1$                    & 159.4 & 6.11 & $0.45\pm0.02$ & $7.81\pm0.89$ & $2.85\pm1.09$ &  $1.23\pm0.05$  & blended\tablefootmark{c} \\     
223422.2629 &     $5_{2,4}-4_{2,3}$ $e_0$                    &  48.4 & 5.15 & $4.77\pm0.35$ & $7.96\pm0.84$ & $2.71\pm1.14$ & $14.29\pm0.50$  & \\     
223616.1420 &     $5_{4,2}-4_{4,1}$, $5_{4,1}-4_{4,0}$ $e_0$ &  95.2 & 2.15\tablefootmark{b} & $2.91\pm0.35$ & $7.70\pm0.84$ & $2.78\pm2.14$ &  $6.95\pm0.23$  & \\  
223691.5380 &     $5_{3,3}-4_{3,2}$ $e_0$                    &  68.2 & 3.96 & $4.81\pm0.35$ & $6.99\pm0.84$ & $2.78\pm1.68$ & $15.87\pm1.76$  & blended\tablefootmark{d} \\     
223697.1880 &     $5_{3,2}-4_{3,1}$ $e_0$                    &  68.2 & 3.96 & $3.04\pm0.35$ & $7.79\pm0.84$ & $1.64\pm1.35$ &  $6.17\pm1.03$  & blended\tablefootmark{e} \\  
225878.2540 &     $3_{1,3}-2_{0,2}$ $o_0$                    &  35.7 & 2.28 & $1.27\pm0.12$ & $7.75\pm0.42$ & $2.67\pm0.44$ &  $3.10\pm0.20$  & \\  

\hline
\multicolumn{9}{c}{dM-2} \\
\hline

105806.4100 & $11_{1,10}-11_{0,11}$ $e_1$                    & 159.4 & 6.11 & $0.38\pm0.02$ & $7.81\pm0.89$ & $2.91\pm1.07$ & $1.02\pm0.05$  & \\     
223422.2629 & $5_{2,4}-4_{2,3}$ $e_0$                        &  48.4 & 5.15 & $2.79\pm0.35$ & $7.96\pm0.84$ & $1.83\pm0.54$ & $5.19\pm0.48$  & \\     
223616.1420 &     $5_{4,2}-4_{4,1}$, $5_{4,1}-4_{4,0}$ $e_0$ &  95.2 & 2.15\tablefootmark{b} & $1.22\pm0.35$ & $7.30\pm0.84$ & $2.84\pm1.33$ & $3.03\pm0.32$  & \\  
223691.5380 &     $5_{3,3}-4_{3,2}$ $e_0$                    &  68.2 & 3.96 & $1.40\pm0.35$ & $7.79\pm0.84$ & $2.80\pm1.68$ & $2.43\pm0.55$  & blended\tablefootmark{d} \\     
223697.1880 &     $5_{3,2}-4_{3,1}$ $e_0$                    &  68.2 & 3.96 & $0.68\pm0.35$ & $7.86\pm0.84$ & $1.22\pm0.97$ & $1.89\pm0.55$  & \\  
225878.2540 &     $3_{1,3}-2_{0,2}$ $o_0$                    &  35.7 & 2.28 & $1.07\pm0.12$ & $7.76\pm0.42$ & $2.52\pm0.42$ & $2.29\pm0.19$  & \\  

\hline                                   %inserts single line
\multicolumn{9}{c}{dM-3} \\
\hline

105806.4100 & $11_{1,10}-11_{0,11}$ $e_1$                    & 159.4 & 6.11 & $0.32\pm0.02$ & $7.89\pm0.89$ & $3.36\pm0.89$ & $1.02\pm0.05$  & \\     
223422.2629 & $5_{2,4}-4_{2,3}$ $e_0$                        &  48.4 & 5.15 & $0.76\pm0.35$ & $7.92\pm0.84$ &  ...          & $2.87\pm1.35$  & \\     
223616.1420 &     $5_{4,2}-4_{4,1}$, $5_{4,1}-4_{4,0}$ $e_0$ &  95.2 & 2.15\tablefootmark{b} & $      <0.35$ & ...           &  ...          & $<1.75$        & \\  
223691.5380 &     $5_{3,3}-4_{3,2}$ $e_0$                    &  68.2 & 3.96 & $1.40\pm0.35$ & $7.76\pm0.84$ & $3.43\pm0.90$ & $4.23\pm0.26$  &  \\     
223697.1880 &     $5_{3,2}-4_{3,1}$ $e_0$                    &  68.2 & 3.96 & $0.93\pm0.35$ & $7.84\pm0.84$ & $4.18\pm1.05$ & $3.43\pm0.31$  & \\  
225878.2540 &     $3_{1,3}-2_{0,2}$ $o_0$                    &  35.7 & 2.28 & $0.76\pm0.12$ & $7.71\pm0.42$ & $0.87\pm0.45$ & $0.71\pm0.11$  & 1st component\\  
            &                                                &       &      & $0.66\pm0.12$ & $9.05\pm0.42$ & $1.22\pm0.45$ & $0.83\pm0.18$  & 2nd component\\

\hline                                   %inserts single line
\end{tabular}
\tablefoot{
\tablefoottext{a}{Line frequencies of the detected lines}
\tablefoottext{b}{Effective line strengths of each single line}
\tablefoottext{c}{This line is blended with the H$^{13}$CCCN $J=12-11$ line at 105799.1130 MHz toward dM-1.}
\tablefoottext{d}{This line is blended with the U223694.8 line.}
\tablefoottext{e}{This line is blended with the U223694.3 line and the acetone line at 223692.1 MHz.}
}

\end{table*}

%\clearpage

\begin{figure*}
\centering
\includegraphics[angle=-90,width=0.75\textwidth]{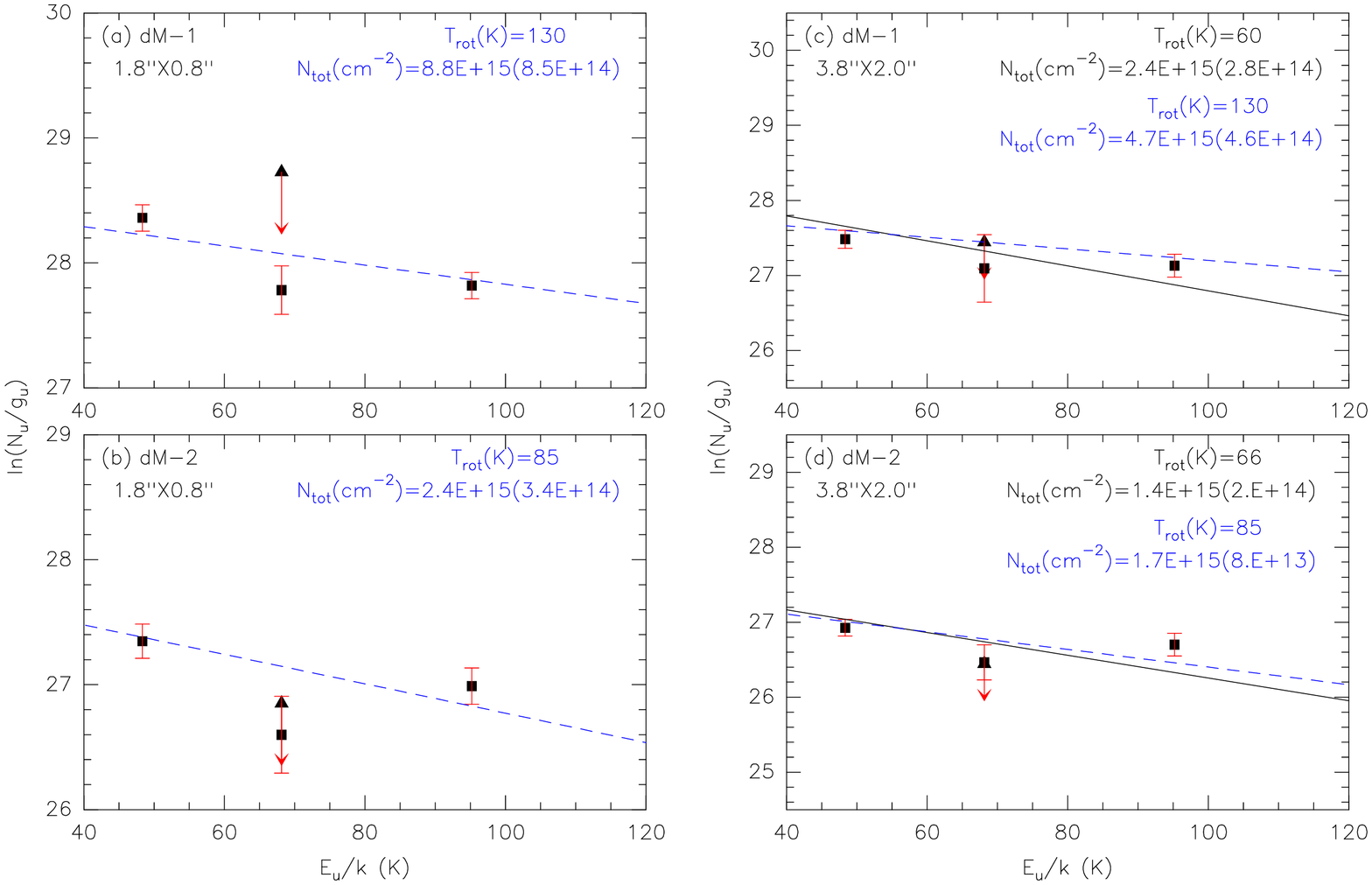}
\caption{\dmeth\ population diagrams of dM-1 and dM-2. (a)--(b) Population diagrams of the high angular resolution data. The rotational temperatures are assumed to be 130 K and 85 K for dM-1 and dM-2, respectively. (c)--(d) Population diagrams for the data smoothed to a resolution of $3\farcs8\times2\farcs0$. Low rotational temperatures and higher ones are assumed to be 60 K and 130 K for dM-1 and 66 K and 85 K for dM-2, respectively. The fits of low temperatures are plotted in black lines and higher temperatures in blue dashed lines. The \dmeth\ column densities are estimated by a least-squares fit, where the uncertainties include the statistical error, the rms noise, and calibration uncertainties ($10\%$).}
\label{Fig-rotation-1}
\end{figure*}

   \begin{figure*}
   \centering
   \includegraphics[angle=-90,width=0.8\textwidth]{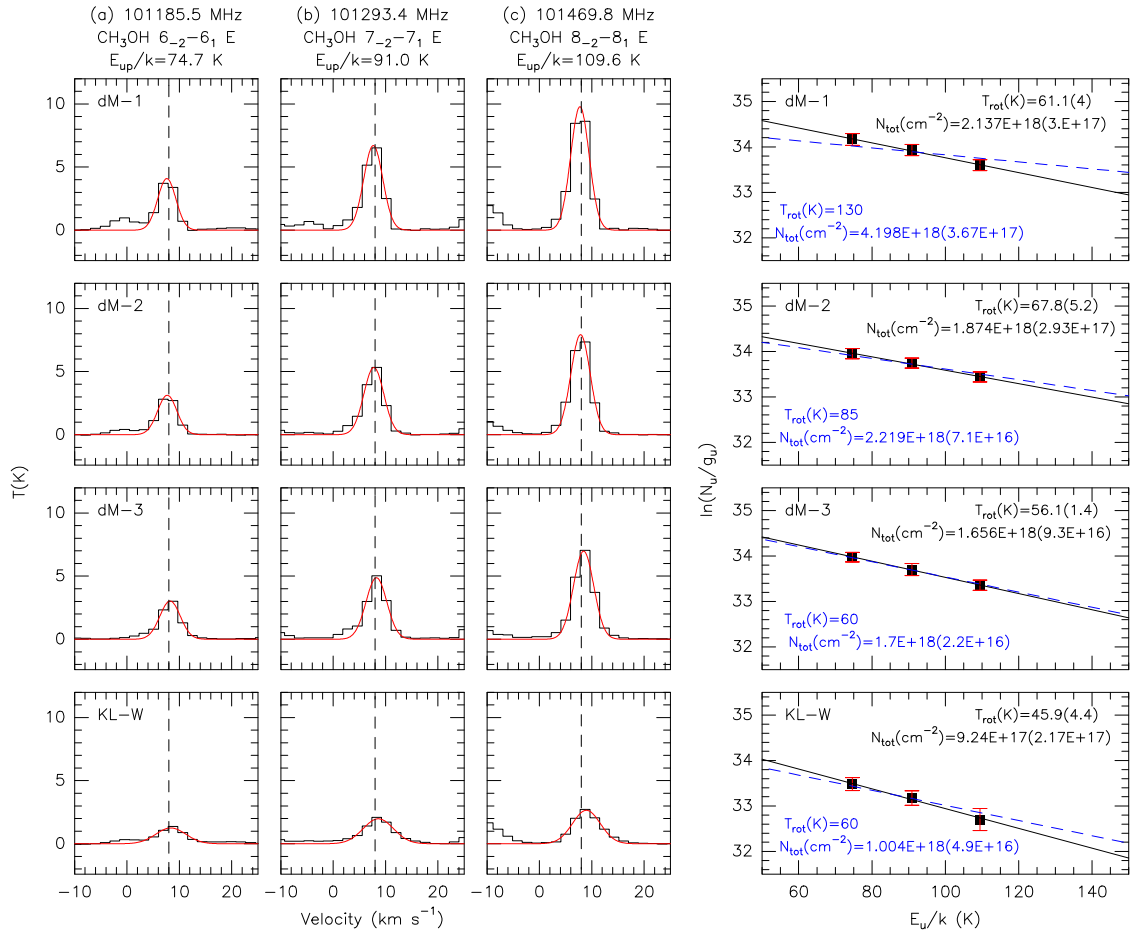}
   \caption{\meth\ spectra ($3\farcs8\times2\farcs0$) toward the four positions identified in Fig. \ref{Fig-CH2DOH-CH3OD-CH3OH-chmap} in the Orion BN/KL region. The methanol rotational temperatures and column densities are estimated by a least-squares fit, where the uncertainties include the statistical error, the rms noise, and calibration uncertainties ($10\%$). Blue dashed lines indicate the fits with fixed temperatures. Black dashed lines indicate a \vlsr\ of 8 \kms.}
              \label{Fig-rotation-2}
    \end{figure*}

       \begin{figure*}
   \centering
   \includegraphics[angle=-90,width=0.8\textwidth]{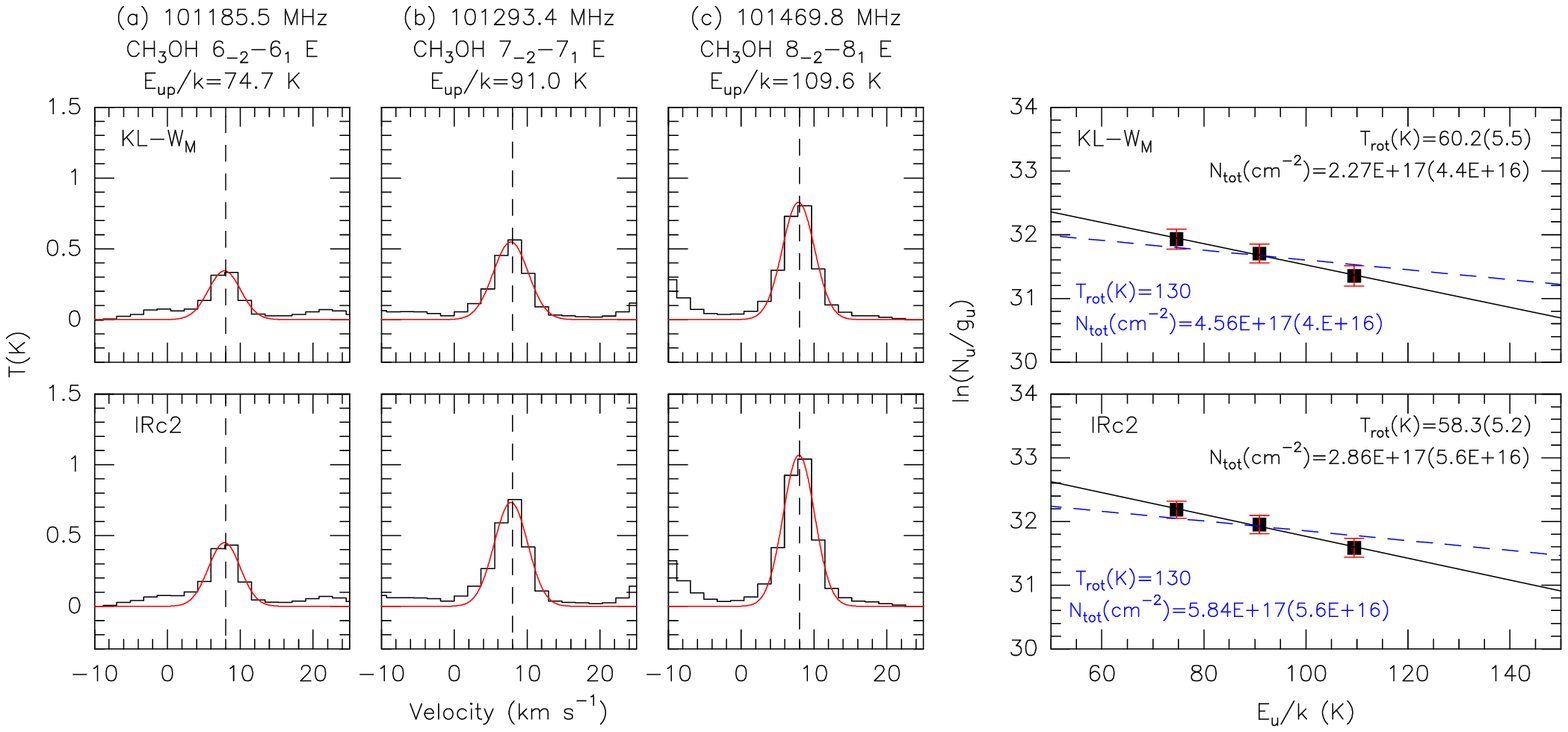}
   \caption{Similar plot as Fig. \ref{Fig-rotation-2}. \meth\ spectra were smoothed to the 30m beam size of 25\arcsec\ to compare with the previous 30m observations by \citet{Menten1988}. Upper panels show the \meth\ spectra and population diagram taken at the same position ($05^{\rm{h}}35^{\rm{m}}14\fs18$, $-05\degr22\arcmin26\farcs5$, J2000) used by \citet{Menten1988} and \citet{Mauersberger1988}, which is located about 5\arcsec\ to the north of KL-W (denoted as KL-W$_{\rm M}$). Lower panels show the \meth\ spectra and population diagram taken at IRc2 ($05^{\rm{h}}35^{\rm{m}}14\fs47$, $-05\degr22\arcmin30\farcs2$, J2000) used by \citet{Jacq1993}. Blue dashed lines indicate the fits with a fixed temperature of 130 K. Black dashed lines indicate a \vlsr\ of 8 \kms. Note that the \meth\ column density derived toward KL-W$_{\rm M}$ is about 20\% lower than that of IRc2.}
              \label{Fig-rotation-3}
    \end{figure*}

           \begin{figure*}
   \centering
   \includegraphics[angle=-90,width=0.85\textwidth]{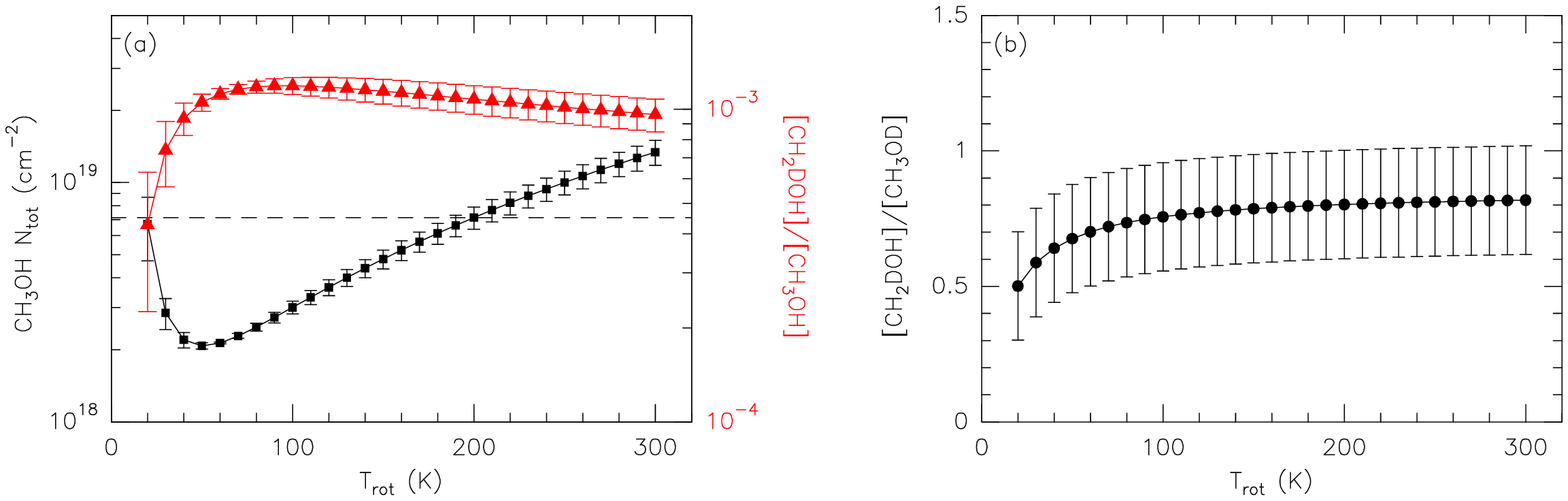}
   \caption{(a) Variation of derived \meth\ column densities with respect to different rotational temperatures toward dM-1. The corresponding [\dmeth]/[\meth] abundance (derived in population diagrams) ratios are shown in red. The best-fit result corresponds to a \meth\ column density of $2.1\times10^{18}$ \cmm\ and a temperature of about 60 K. For temperatures lower than 200 K, the differences in the derived column densities are clearly smaller than a factor of about 3. The [\dmeth]/[\meth] abundance ratios derived from different temperatures are lower than $1.1\times10^{-3}$. (b) Variation of the relative abundance (based on the ratio of two transitions of \dmeth\ and \methd\ as seen in Fig. \ref{Fig-overlay-2} e) of the two isotopologs toward dM-1 as a function of the adopted temperatures. Note that this ratio does not depend much on the temperature.}
              \label{Fig-temp-ntot}
    \end{figure*}

\begin{table*}
\caption{PdBI \meth\ line parameters at dM-1, dM-2, dM-3, and KL-W}             % title of Table
\label{table-CH3OH}      % is used to refer this table in the text
\centering                          % used for centering table
\begin{tabular}{cccccccc}        % centered columns (4 columns)
\hline\hline                 % inserts double horizontal lines
Frequency  & Transition  & $E_{\rm up}/k$  & $S\mu^2$ & $T_{\rm peak}$  & $V_{\rm LSR}$ & $\Delta V$    & $\int TdV$ \\    % table heading 
(MHz)      & ($J_{k}$)   & (K)           & (D$^2$)  &  (K)         & (km s$^{-1}$) & (km s$^{-1}$) & (K km s$^{-1}$)      \\
       
\hline                        % inserts single horizontal line

\multicolumn{8}{c}{dM-1 ($05^{\rm{h}}35^{\rm{m}}14\fs442$, $-05\degr22\arcmin34\farcs86$)} \\
\hline

101185.4530 & $6_{-2}-6_{1}$ E &  74.7 & 0.021 &  $3.72\pm0.05$ & $7.01\pm0.93$ & $4.09\pm1.58$ &  $17.92\pm5.56$  \\     
101293.4150 & $7_{-2}-7_{1}$ E &  91.0 & 0.046 &  $6.50\pm0.05$ & $8.31\pm0.93$ & $4.29\pm0.95$ &  $30.72\pm5.61$  \\     
101469.8050 & $8_{-2}-8_{1}$ E & 109.6 & 0.091 &  $8.59\pm0.05$ & $7.84\pm0.93$ & $4.18\pm0.65$ &  $43.49\pm5.45$  \\  

\hline
\multicolumn{8}{c}{dM-2 ($05^{\rm{h}}35^{\rm{m}}14\fs320$, $-05\degr22\arcmin37\farcs23$)} \\
\hline

101185.4530 & $6_{-2}-6_{1}$ E &  74.7 & 0.021 &  $2.70\pm0.05$ & $7.86\pm0.93$ & $4.33\pm1.98$ &  $14.48\pm5.38$  \\     
101293.4150 & $7_{-2}-7_{1}$ E &  91.0 & 0.046 &  $5.29\pm0.05$ & $8.26\pm0.93$ & $4.48\pm1.19$ &  $25.49\pm5.52$  \\     
101469.8050 & $8_{-2}-8_{1}$ E & 109.6 & 0.091 &  $7.36\pm0.05$ & $8.76\pm0.93$ & $4.40\pm0.75$ &  $37.07\pm5.34$  \\  

\hline

\multicolumn{8}{c}{dM-3 ($05^{\rm{h}}35^{\rm{m}}14\fs107$, $-05\degr22\arcmin37\farcs43$)} \\

\hline
101185.4530 & $6_{-2}-6_{1}$ E &  74.7 & 0.021 &  $2.98\pm0.05$ & $8.81\pm0.93$ & $4.46\pm1.82$ &  $14.28\pm4.65$  \\     
101293.4150 & $7_{-2}-7_{1}$ E &  91.0 & 0.046 &  $5.00\pm0.05$ & $8.31\pm0.93$ & $4.59\pm1.14$ &  $23.78\pm4.76$  \\     
101469.8050 & $8_{-2}-8_{1}$ E & 109.6 & 0.091 &  $6.84\pm0.05$ & $8.76\pm0.93$ & $4.54\pm0.74$ &  $33.54\pm4.62$  \\  

\hline

\multicolumn{8}{c}{KL-W ($05^{\rm{h}}35^{\rm{m}}14\fs159$, $-05\degr22\arcmin28\farcs25$)} \\

\hline
101185.4530 & $6_{-2}-6_{1}$ E &  74.7 & 0.021 &  $2.70\pm0.05$ & $8.76\pm0.93$ & $6.52\pm6.13$ &   $8.65\pm6.05$  \\     
101293.4150 & $7_{-2}-7_{1}$ E &  91.0 & 0.046 &  $4.16\pm0.05$ & $8.31\pm0.93$ & $6.72\pm3.44$ &  $14.09\pm5.65$  \\     
101469.8050 & $8_{-2}-8_{1}$ E & 109.6 & 0.091 &  $5.40\pm0.05$ & $8.76\pm0.93$ & $6.17\pm2.24$ &  $17.38\pm5.25$  \\

\hline                                   %inserts single line
\end{tabular}

\tablefoot
{The R.A. and Dec. coordinates at the J2000.0 epoch are given for the four sources. Frequency, $E_{\rm up}/k$, and $S\mu^2$ data were taken from the JPL database. The \vlsr\ is measured at the peak temperatures, and the line widths and integrated intensities are estimated by fitting a Gaussian profile.
%\tablefootmark{a}{Frequency, $E_{\rm up}/k$, and $S\mu^2$ data were taken from the JPL database.}
%\tablefoottext{b}{The \vlsr\ is measured at the peak temperatures, and the line widths and integrated intensities are estimated by fitting a Gaussian profile.}
}

\end{table*}

\begin{table*}
\caption{PdBI \methd\ $5_{-1}-4_{-1}$ E line parameters at dM-1, dM-2, and dM-3}             % title of Table
\label{table-CH3OD}      % is used to refer this table in the text
\centering                          % used for centering table
\begin{tabular}{cccccc}        % centered columns (4 columns)
\hline\hline                 % inserts double horizontal lines
Position  &  $T_{\rm peak}$  & $V_{\rm LSR}$ & $\Delta V$    & $\int TdV$       &  Comment \\ % table heading 
        &   (K)            & (km s$^{-1}$) & (km s$^{-1}$) & (K km s$^{-1}$)  &    \\
       
\hline                        % inserts single horizontal line

dM-1 &  $1.91\pm0.24$ & $7.69\pm0.42$ & $2.19\pm0.42$ &  $4.47\pm0.27$  & \\     
dM-2 &  $1.82\pm0.22$ & $7.58\pm0.42$ & $2.87\pm0.42$ &  $5.55\pm0.36$  & \\     
dM-3 &  $1.60\pm0.30$ & $7.43\pm0.42$ & $1.39\pm0.42$ &  $2.36\pm0.21$  & 1st component\\  
     &  $0.83\pm0.27$ & $9.28\pm0.42$ & $1.94\pm0.41$ &  $1.71\pm0.28$  & 2nd component\\

\hline                                   %inserts single line
\end{tabular}

%\tablefoot
%{
%The line parameters of the \methd\ $5_{-1}-4_{-1}$ E line at 226185.9 MHz are estimated by fitting a Gaussian profile.
%\tablefootmark{a}{Frequency, $E_{\rm up}/k$, and $S\mu^2$ data were taken from the JPL database.}
%\tablefoottext{b}{The \vlsr\ is measured at the peak temperatures, and the line widths and integrated intensities are estimated by fitting a Gaussian profile.}
%}

\end{table*}

\end{appendix}

%
%---------------------------------------------------------------------------
%

\end{document}